\documentstyle[aps,preprint,psfig,axodraw]{revtex}
\tightenlines
\begin{document}
\draft
\title{PERIPHERAL NN-SCATTERING: ROLE OF \\ DELTA-EXCITATION,
CORRELATED TWO-PION \\ AND VECTOR MESON EXCHANGE\footnote{Work supported in 
part by BMBF.}}
\author{N. Kaiser, S. Gerstend\"orfer and W. Weise}
\address{Physik-Department, Technische Universit\"{a}t M\"{u}nchen,\\
    D-85747 Garching, Germany} 

\bigskip

\bigskip
\maketitle
\begin{abstract}
We evaluate, within one-loop chiral perturbation theory, the two-pion exchange 
diagrams with single and double delta-isobar excitation contributing to 
elastic NN-scattering. We find that virtual $\Delta$-excitation processes (in 
the static limit) produce the correct amount of isoscalar central attraction as
needed in the peripheral partial waves with $L\geq 3$. Furthermore we compute 
the two-loop diagrams involving the $\pi\pi$-interaction (so-called correlated 
$2\pi$-exchange). Contrary to common believe these processes lead to negligibly
small and repulsive corrections to the NN-potential. The exchange of vector
mesons ($\rho,\,\omega$) turns out to be important for the F-wave phase shifts
above $T_{lab}=150$ MeV. Without adjustable parameters we are able to reproduce
the empirical NN phase shifts up to 350 MeV for $L\geq 3$ and up to about
(50--80) MeV for the D-waves. This is therefore the characteristic window in
which the NN-interaction is basically governed by chiral symmetry. Not 
surprisingly, the lower partial waves require non-perturbative methods and
additional short-distance parametrizations of the NN-dynamics. 
\end{abstract}

\vskip 5cm

\newpage

\section{INTRODUCTION AND SUMMARY}
In a recent paper \cite{kbw} we have developed a framework for the application
of chiral perturbation theory to low energy elastic nucleon-nucleon scattering.
In that framework a systematic expansion of the NN T-matrix in powers of small
external momenta and quark masses is performed by evaluating tree and loop
diagrams with vertices taken from an effective chiral Lagrangian. The latter is
an efficient tool to implement the (chiral) symmetry constraints on the
dynamics of pions (the Goldstone bosons of spontaneous chiral symmetry breaking
in QCD). Since such an approach to NN-scattering focuses on the dynamics of 
pions it is expected to work within the kinematical domain where the NN force 
is dominated by one- and two-pion exchange. These are the peripheral NN partial
waves below the inelastic $NN\pi$-threshold.

In ref.\cite{kbw} the $2\pi$-exchange contributions were worked out up to third
order in small momenta including one-loop graphs with a vertex from the second
order chiral $\pi N$-Lagrangian ${\cal L}^{(2)}_{\pi N}$. This part of the
Lagrangian involves several additional low-energy constants $c_i,\,(i=1,2,3,4)
$, which have been recently determined from a fit to many low energy
pion-nucleon data \cite{pin}. Since some of these constants $c_i$ are much
larger than the natural scale $1/2M$ ($M$ being the nucleon mass) they indeed 
produce the major $2\pi$-exchange effect in  NN-scattering. In particular, the 
isoscalar central potential is almost entirely given by the constant $c_3$ 
proportional to the so-called nucleon axial polarizability \cite{pin,axpola}. 
As shown in ref.\cite{pin} it is the $\Delta(1232)$-resonance that makes the 
dominant contribution to the low energy  constants $c_{2,3,4}$ if one
interprets their values in terms of resonance exchange. Therefore the  
$2\pi$-exchange effects proportional to $c_{3,4}$ as calculated in 
ref.\cite{kbw} can be approximately identified as single $\Delta$-excitation 
graphs for which one has ignored the energy dependence of the 
$\Delta$-propagator by using just a contact $\pi\pi NN$-vertex. However, at
the inelastic $NN\pi$-threshold the nucleon kinetic energy $T_{lab}$ becomes 
comparable to the $\Delta N$-mass splitting of 293 MeV, and a treatment of 
important $\Delta$-dynamics via contact interactions may be too crude. This is
also indicated by the results in ref.\cite{kbw} which show too large attraction
in the F-waves above $T_{lab}= 180$ MeV and in the D-waves even for smaller
energies around 50 MeV. One can expect that explicit $\Delta(1232)$ degrees of
freedom cure this problem (at least partly).   

In the past the $\Delta$-excitation processes in NN-scattering have been 
considered in various  approaches, either via dispersion theoretical methods
\cite{disp} or using $N\Delta$-transition potentials \cite{durso,pand,green} 
which neglect some (technically complicated) parts of the diagrams. At present
no proper quantum field theoretical evaluation of these diagrams exists, which
would avoid artificial "form factors" or cutoffs. As in ref.\cite{bonn} such 
"form factors"are often introduced in order to enforce convergent loop
integrals. In this  work we evaluate the single and double $\Delta$-excitation 
graphs in the static limit $M\to \infty $ (and also their first relativistic 
$1/M$-correction) using covariant perturbation theory and dimensional 
regularization. In one-loop order to which we are working here the 
divergences of the diagrams show up only as purely polynomial NN-amplitudes 
which do not contribute to the phase shifts with orbital angular
momentum $L\geq 2$ and the mixing angles with $J\geq2$. Therefore we can study
the effects due to $\Delta$-excitation in a completely parameterfree fashion. 

Next we consider, at two-loop order, the two-pion exchange diagrams involving 
the chiral $\pi\pi$-interaction, so-called correlated $2\pi$-exchange. In 
phenomenological approaches \cite{disp,durso} such processes are often 
identified with the enhancement around masses of 550 MeV observed in the 
isoscalar central $(\pi\pi \to N \overline N)$ spectral function proportional
to $|f_{0+}|^2$. Such a behavior is found in the dispersion theoretical 
analyses of the $\pi N$-scattering data \cite{hoehler} and often called 
"$\sigma$-meson". After transforming the respective NN-amplitudes into a local 
coordinate space potential, we find that these two-loop diagrams involving the 
chiral $\pi\pi$-interaction lead to rather different effects. In particular the
resulting isoscalar central potential turns out to be weakly repulsive, 
contrary to common believe. The repulsive nature of isoscalar central potential
finds its explanation in the isospin-zero S-wave $\pi\pi$-amplitude which 
becomes repulsive sufficiently far below the threshold. In any case, all 
effects due to correlated $2\pi$-exchange (at two-loop) that we find here in 
(heavy baryon) chiral perturbation theory are in fact negligibly small. This  
means that the enhancement showing up in the spectral function $|f_{0+}|^2$ 
cannot simply be identified with the
(perturbative two-loop) correlated $2\pi$-exchange  diagram.

With decreasing orbital angular momentum $L$ we find that the 
description of the empirical F-wave NN phase shifts above $T_{lab}=150$ MeV 
requires a further (well-known) ingredient, namely the exchange of vector 
mesons, $\rho$ and $\omega$. While $\omega$-meson exchange (using $g_{\omega N}
^2/4\pi= 12.9$ for its coupling constant) provides the necessary overall 
repulsion, the $\rho$-meson with its large tensor-to-vector coupling ratio
($\kappa_\rho=6$)  leads to the correct splitting of the singlet and the three
triplet F-waves.  

In D-waves we find deviations from the empirical NN phase shifts already at 
$T_{lab}=100$ MeV (partly even for lower $T_{lab}$). For these low angular 
momentum partial waves the van der 
Waals behavior of the $2\pi$-exchange NN-potential becomes problematic, since
its $r^{-6}$-singularity practically extinguishes the natural centrifugal 
barrier effects coming from the wave function $r^2 j_2(pr)^2$ (with $j_2(pr)$ 
a spherical Bessel function). At such kinematics the short range NN-repulsion
starts to become essential. Its dynamical origin lies of course outside the
perturbative interaction of point-like baryons and mesons treated here. For 
some recent attempts on this problem in the framework of effective field 
theory see ref.\cite{kolck,savage}

We can summarize our work as follows: Within systematic perturbation theory
based on chiral symmetry one finds an accurate description of the empirical NN
phase shifts in the partial waves with $L\geq 3$ up to  $T_{lab}=350$ MeV and 
partly up to 80 MeV in D-waves with the following ingredients:
\begin{itemize} 
\item[1)]  Point-like one-pion exchange (without introducing an unmeasurable 
$\pi N$ form factor) 
\item[2)] Iterated one-pion exchange 
\item[3)] Irreducible two-pion exchange with only nucleons in intermediate 
states (plus their first relativistic $1/M$-correction)
\item[4)] Two-pion exchange with single and double $\Delta$-isobar excitation 
(in the static limit)
\item[5)] Vector meson ($\rho$ and $\omega$) exchange with standard values of 
the coupling constants but no ad hoc form factors
\end{itemize}

These components comprise the dynamics of the peripheral nucleon-nucleon 
interaction. The emerging physical picture is of course not entirely new.  
However, our  calculation is based on an effective chiral Lagrangian and we 
apply rigorous methods of perturbative quantum field theory (covariant Feynman
graphs and dimensional regularization). Thus there is no need to introduce 
extra cut offs or ad hoc form factors which sometimes obscure the real
physics. 

On the experimental side there are upcoming precision data from the Indiana 
Cooler Synchrotron Facility (IUCF), which will lead to an improved NN phase 
shift analysis in the energy range below the pion production threshold. We 
propose to use the chiral NN phase shifts with $L\geq 3$ presented here as 
input in a future phase shift analysis.

\section{BASIC FORMALISM}

In this section, we briefly review some basic formalism needed to describe 
elastic nucleon-nucleon scattering. In the center of mass frame the on-shell 
T-matrix for the process $N(\vec p\,) + N(-\vec p\,) \to  N(\vec p\,')+N(-\vec 
p\,')$ takes the following general form,
\begin{eqnarray} {\cal T}_{NN} &=& V_C+ \vec \tau_1 \cdot \vec \tau_2  W_C +
\big[V_{S}+ \vec \tau_1 \cdot \vec \tau_2  W_{S} \big]\,\vec\sigma_1
\cdot \vec \sigma_2+ \big[ V_T + \vec \tau_1 \cdot \vec \tau_2 W_T 
\big]\,  \vec  \sigma_1 \cdot \vec q \, \vec \sigma_2 \cdot \vec q  \nonumber 
\\  & &+ \big[ V_{SO}+\vec\tau_1 \cdot \vec \tau_2 W_{SO} \big] \,i( \vec 
\sigma_1 +\vec  \sigma_2)\cdot (\vec q\times \vec p\,) \nonumber \\ & & + \big[
V_Q+\vec \tau_1 \cdot \vec \tau_2 W_Q \big]\,\vec\sigma_1\cdot(\vec q\times 
\vec p\,) \,\vec \sigma_2 \cdot(\vec q\times \vec p\,)\,.\end{eqnarray}
The ten complex functions $V_C, \dots, W_Q$ depend on the center of mass 
momentum  $p= |\vec p\,| = |\vec p\,'\,|$ and the momentum transfer $q = |\vec
q\,|$ with $\vec q = \vec p\,' - \vec p$. The subscripts refer to the central,
spin-spin, tensor, spin-orbit and quadratic spin-orbit components, each of 
which occurs in an  isoscalar $(V)$ and an isovector $(W)$ version. In terms
of the nucleon laboratory kinetic energy $T_{lab}$ the center of mass momentum
$p$ is given as $p=\sqrt{T_{lab}M/2}$. 

In order to compute the NN phase shifts and mixing angles one needs the matrix
elements  of ${\cal T}_{NN}$ in the  $LSJ$-basis, where $L=J-S,J,J+S,\,S=0,1$ 
and $J=0,1,2,\dots$ denote the orbital angular momentum, the total spin and the
total angular momentum respectively. The explicit form of the partial wave
projection formulas can be found in section 3 of ref.\cite{kbw}. Phase shifts
and mixing angles are then given  perturbatively as 
\begin{eqnarray} \delta_{LSJ} &=& {M^2 p \over 4\pi E}\, {\rm Re}\,\langle LSJ 
|{\cal T}_{NN} | L S J\,\rangle \,, \\ \epsilon_J &=& {M^2p\over 4\pi
E }\,{\rm Re}\,\langle J-1,1J|{\cal T}_{NN}|J+1,1J\,\rangle \,,  \end{eqnarray}
with the center of mass nucleon energy $E=\sqrt{M^2+p^2}$. The perturbative 
expressions in eqs.(2,3) apply only if the phase shifts $\delta_{LSJ}$ and 
mixing angles $\epsilon_J$ are sufficiently small in order not to have any
substantial violations of unitarity. We calculate reliably only those phase
shifts and mixing angles which are smaller than $ 10^\circ$ in magnitude. For
these cases perturbation theory is well justified.

\section{ONE LOOP DIAGRAMS WITH DELTA-EXCITATION}
The one-loop diagrams with single and double $\Delta$-isobar excitation are
shown in Fig.1. The first diagram of triangle shape is specific for
a calculation based on a chiral effective  $\pi N$-Lagrangian, since it
involves the Weinberg-Tomozawa isovector $NN\pi\pi$-contact vertex. Since all 
Feynman rules for the nucleon vertices and propagator are well documented in
the recent review \cite{review}, we give here only some details associated with
the $\Delta$-isobar. In the heavy mass limit \cite{hemm} where one considers 
the external momenta and the delta-nucleon mass splitting $\Delta= M_\Delta- M=
293$ MeV small compared to the nucleon mass, $M=939$ MeV, one gets for the 
delta-propagator and the $\Delta\to  \pi^a N$ transition vertex
\begin{equation} {i \over k_0 - \Delta +i0^+} \,\,, \quad\qquad \qquad - {3 g_A
\over 2 \sqrt2 f_\pi} \vec S \cdot \vec l \,\, T_a \,\,. \end{equation}
Here $k^\mu$ denotes the four-momentum of the propagating delta (with its 
energy $k_0$ counted modulo the large nucleon mass $M$) and $l^\mu$ is the 
four-momentum of the emitted pion with isospin $a$. The $2\times 4$ spin and
isospin transition matrices $S_i$ and $T_a$ satisfy the relations $S_i\,
S_j^\dagger =(2\delta_{ij} - i \epsilon_{ijk}\sigma_k )/3$ and $T_a \, 
T_b^\dagger =(2\delta_{ab} - i \epsilon_{abc} \tau_c )/3$ \cite{erwe}. For the
$\pi N\Delta$ coupling constant we have already inserted the large-$N_c$ value 
$g_{\pi N\Delta} = 3 g_{\pi N}/\sqrt{2}$ together with the Goldberger-Treiman
relation $g_{\pi N} = g_A M/f_\pi =13.4$ (with $f_\pi =92.4 $ MeV this gives
$g_A=1.32$). For the $\Delta \to \pi N$  decay width one finds in this case
\begin{equation} \Gamma(\Delta \to \pi N) = {3g_A^2 \over 32 \pi f_\pi^2} 
{E_N +M \over M+\Delta} (E_N^2-M^2)^{3/2} = 110.6 \,{\rm MeV}\,, \end{equation}
a number which is in good agreement with the empirical decay width
$\Gamma(\Delta \to \pi N) = (115 \pm 5)$ MeV. Here $E_N = M+(\Delta^2-m_\pi^2)
/ 2M_\Delta=966$\, MeV denotes the center of mass energy of the decay nucleon. 
We note that eq.(5) for the $\Delta$-decay width derives from a relativistic
calculation using Rarita-Schwinger spinors \cite{peccei}. Since the momentum of
the decay  pion is not so small, namely $227.3$ MeV, one would loose important 
kinematical factors in a non-relativistic approximation and overestimate the
$\Delta$-decay width \cite{erwe}.    

\begin{center}
\SetScale{0.8}
\SetWidth{1.5}
  \begin{picture}(364,72)
\Line(0,0)(0,85)
\Line(57,0)(57,85)
\Line(4,20)(4,66)
\BCirc(2,20){2}
\BCirc(2,66){2}
\DashLine(4,20)(57,42.5){6}
\DashLine(4,66)(57,42.5){6}

\Line(99,0)(99,85)
\Line(156,0)(156,85)
\Line(103,20)(103,66)
\BCirc(101,20){2}
\BCirc(101,66){2}
\DashLine(103,20)(156,20){6}
\DashLine(103,66)(156,66){6}

\Line(199,0)(199,85)
\Line(256,0)(256,85)
\Line(252,20)(252,66)
\DashLine(199,20)(252,66){6}
\DashLine(199,66)(252,20){6}
\BCirc(254,66){2}
\BCirc(254,20){2}

\Line(299,0)(299,85)
\Line(356,0)(356,85)
\Line(303,20)(303,66)
\Line(352,20)(352,66)
\DashLine(303,20)(352,20){6}
\DashLine(303,66)(352,66){6}
\BCirc(301,20){2}
\BCirc(301,66){2}
\BCirc(354,20){2}
\BCirc(354,66){2}

\Line(398,0)(398,85)
\Line(455,0)(455,85)
\Line(402,20)(402,66)
\Line(451,20)(451,66)
\BCirc(400,20){2}
\BCirc(400,66){2}
\BCirc(453,20){2}
\BCirc(453,66){2}
\DashLine(402,20)(451,66){6}
\DashLine(402,66)(451,20){6}

  \end{picture}

\medskip

{\it Fig.1: One-loop $2\pi$-exchange diagrams with single and double 
$\Delta(1232)$-excitation }

\end{center}

\medskip

Now we give the one-loop NN-amplitudes which result from evaluating the
diagrams shown in Fig.1 in the heavy mass limit, $M\to \infty$. For the sake of
simplicity we omit purely polynomial terms which do not contribute to the phase
shifts with $L\geq 2$ and to mixing angles with $J\geq 2$ (see section 4.1 in
ref.\cite{kbw}). The divergences of the loop diagrams are actually included 
in such (irrelevant) polynomial terms. We note that all one-loop vertex and
self energy corrections with $\Delta$-isobar excitation to the $1\pi$-exchange
diagram lead only to mass and coupling constant renormalization; they do not
introduce a pion-nucleon "form factor" (see section 4.1 in \cite{kbw}). We find
the following analytical results for the three classes of diagrams shown in
Fig.1: 

\medskip

a) $\Delta$-excitation in triangle graphs: 
\begin{equation} W_C= {g_A^2 \over 192 \pi^2 f_\pi^4} \Big\{ (6\Sigma-w^2)L(q)
+12\Delta^2 \Sigma\,  D(q)\Big\}\,\,.\end{equation} 

b) Single $\Delta$-excitation in box graphs: 
\begin{eqnarray}V_C&=& {3g_A^4 \over 32\pi f_\pi^4 \Delta}(2m_\pi^2+q^2)^2
A(q)  \,\,, \\ W_C&=& {g_A^4 \over 192 \pi^2 f_\pi^4}\Big\{(12\Delta^2 -20
m_\pi^2-11 q^2)L(q)+6 \Sigma^2 \,D(q)\Big\}\,\,, \\ V_T&=& 
-{1\over q^2}\,V_S={3g_A^4 \over 128 \pi^2f_\pi^4}\Big\{-2 L(q)+(w^2-4
\Delta^2) D(q)\Big\}\,\,, \\ W_T&=&-{1\over q^2}\, W_S= {g_A^4 \over
128\pi f_\pi^4\Delta} w^2A(q)\,\,.\end{eqnarray}  

c) Double $\Delta$-excitation in box graphs: 
\begin{eqnarray}V_C&=& {3g_A^4 \over 64\pi^2 f_\pi^4} \Big\{-4\Delta^2
L(q)+ \Sigma\, \big[H(q)+(\Sigma+8\Delta^2)  D(q)\big] \Big\}\,\,, \\  W_C&=&
{g_A^4 \over  384\pi^2 f_\pi^4} \Big\{(12\Sigma-w^2)L(q) +3\Sigma\, \big[H(q)+ 
(8  \Delta^2-\Sigma) D(q) \big] \Big\}\,\,,\\  V_T&=&- 
{1\over q^2}\,V_S={3g_A^4\over512\pi^2f_\pi^4} \Big\{ 6 L(q)+ (12 \Delta^2-
w^2) D(q)\Big\}\,\,, \\ W_T&=& -{1\over q^2}\,W_S={g_A^4 \over 1024
\pi^2 f_\pi^4} \Big\{ 2L(q)+(4 \Delta^2+w^2) D(q) \Big\}  \,\,.
\end{eqnarray}   

The following set of loop functions and abbreviations has been used to express
these NN-amplitudes,
\begin{eqnarray} L(q)&=& {w\over q} \ln{w+q\over 2m_\pi}\,\,, \qquad w=\sqrt{
4m_\pi^2 +q^2}\,\,,\\ A(q) &=& {1\over 2q} \arctan{q\over 2m_\pi} \,\,, \\ D(q)
&=& {1\over \Delta}\int_{2m_\pi}^\infty {d\mu \over \mu^2 +q^2} \arctan{\sqrt
{\mu^2-4m_\pi^2} \over 2\Delta}\,\,, \\ H(q)&=&{2\,\Sigma \over w^2-4 \Delta^2}
\Big[L(q)-L(2\sqrt{\Delta^2-m_\pi^2})\Big]\,\,, \\ \Sigma &=& 2m_\pi^2+q^2-2
\Delta^2 \,\,.  \end{eqnarray}  

The isoscalar central $V_C$ and isovector tensor amplitude $W_T$ coming from 
the box graphs with single $\Delta$-excitation in eqs.(7,10) show an 
interesting feature. The dependence on the delta-nucleon mass splitting 
$\Delta$ is a trivial factor $\Delta^{-1}$. This means that, in the sum of the 
corresponding planar and crossed box graphs, the energy dependence of the 
delta-propagator has effectively disappeared. A $\pi\pi NN$-contact vertex 
($-c_3=2c_4 = g_A^2/2\Delta$) proportional to the inverse mass splitting
$\Delta^{-1}$ instead of the propagating $\Delta$-isobars would give exactly 
the same result. Another way to understand this coincidence is to compute the 
sum of the energy denominators for  all possible time-orderings which gives 
\begin{equation} {2\omega_1 \omega_2(\omega_1+\omega_2) +\Delta(\omega_1^2+3
\omega_1\omega_2+\omega^2_2) +\Delta^2(\omega_1+\omega_2) \over   \Delta 
\omega_1\omega_2(\omega_1+\omega_2)(\omega_1+\Delta)(\omega_2+\Delta) }
\end{equation} 
for the planar box graph and 
\begin{equation} {\omega_1^2+\omega_1\omega_2+\omega^2_2 +\Delta(\omega_1+
\omega_2) \over  \omega_1\omega_2(\omega_1+\omega_2)(\omega_1+\Delta)
(\omega_2+\Delta) } \end{equation}
for the crossed box graph, with $\omega_1$ and $\omega_2$ denoting the on-shell
energies of the two exchanged pions. The sum of the two expressions in eq.(20) 
and eq.(21) leads to the surprisingly simple expression $2(\Delta \omega_1
\omega_2)^{-1}$ in which the mass splitting $\Delta$ has factored out. In all 
other cases there is however a nontrivial dependence of the NN-amplitudes on 
the mass splitting $\Delta$, in particular due to the occurrence of the loop 
function $D(q)$ for which we have given its spectral representation in eq.(17).

The NN-amplitudes in momentum space given above can be transformed into 
local coordinate space potentials (disregarding zero-range $\delta^3(\vec r\,)
$-terms) in the form of continuous superpositions of Yukawa functions (for 
details see section 6 in ref.\cite{kbw}). The mass spectra entering in this 
representation are given by imaginary parts of the NN-amplitudes analytically 
continued to time-like momentum transfer $q=i\mu-0^+$.  This requires the 
knowledge of the imaginary parts of the loop functions for $\mu > 2m_\pi$,
\begin{eqnarray} L(i\mu)&=& {\sqrt{ \mu^2-4m_\pi^2}\over \mu}\bigg[ \ln{\mu + 
\sqrt{\mu^2-4m_\pi^2} \over 2m_\pi} -i \,{\pi\over 2} \bigg]\,\,,\\ 
A(i\mu) &=& {1\over 4\mu} \bigg[ \ln{\mu+2m_\pi \over
\mu-2m_\pi } + i\, \pi \bigg] \,\,, \\ {\rm Im}\, D(i\mu) &=& {\pi\over 2 \mu\,
\Delta}  \arctan{\sqrt{\mu^2-4m_\pi^2} \over 2\Delta}\,\,. \end{eqnarray}  
Here we have also given the real parts of $L(i\mu)$ and $A(i\mu)$ which
will be needed later in the discussion of the correlated two-pion exchange.
As in our previous work \cite{kbw} we will denote $r$-space potentials by a 
tilde, i.e. $\widetilde V_C(r)$ is the isoscalar central potential, etc. 

The attractive isoscalar central potential $\widetilde V_C(r)$ generated 
by the $\Delta$-excitation diagrams is shown in Fig.2. The main contribution
comes  from the four graphs with single $\Delta$-excitation  and it has the 
following  simple analytical form 
\begin{eqnarray} \widetilde V_C^{(N\Delta)}(r) &=& -{3 g_A^4 \over 64\pi^2
f_\pi^4 \Delta} {e^{-2x} \over r^6} (6+12x+10x^2+4x^3+x^4) \nonumber \\ &=& -
{36\over \Delta} \Big\{2 \big[\widetilde W_T^{(1\pi)}(r)\big]^2 +
\big[\widetilde W_S^{(1\pi)}(r)\big]^2 \Big\}  \,\,, \end{eqnarray}
with the abbreviation $x= m_\pi r$. We have expressed this potential in the 
second line in terms of squares of the $1\pi$-exchange isovector spin-spin and
tensor potentials  $\widetilde W_{S,T}^{(1\pi)}(r)$.  For the isovector 
spin-spin and tensor potentials coming from the
single $\Delta$-excitation graphs one finds, similarly,
\begin{eqnarray} \widetilde W_S^{(N\Delta)}(r) &=& {g_A^4 \over 192\pi^2
f_\pi^4 \Delta} {e^{-2x} \over r^6} (1+x)(3+3x+2x^2) \nonumber \\ &=& 
{4\over \Delta} \Big\{\big[\widetilde W_T^{(1\pi)}(r)\big]^2 -
\big[\widetilde W_S^{(1\pi)}(r)\big]^2 \Big\} \,\,, \\
\widetilde W_T^{(N\Delta)}(r) &=& -{g_A^4 \over 192\pi^2
f_\pi^4 \Delta} {e^{-2x} \over r^6} (1+x)(3+3x+x^2) \nonumber \\ &=& 
{4\over \Delta} \widetilde W_T^{(1\pi)}(r) \Big\{\widetilde W_S^{(1\pi)}(r)
-\widetilde W_T^{(1\pi)}(r)\Big\}  \,\,. \end{eqnarray}
The relative weights between the $1\pi$-exchange spin-spin and tensor potential
in eqs.(25,26,27) can be easily understood from relations between the tensor
operator $S_{12}(\hat r)= 3\vec\sigma \cdot \hat r\, \vec \sigma_2 \cdot \hat r
- \vec \sigma_1\cdot \vec \sigma_2$ and the spin-spin operator $\vec \sigma_1
\cdot \vec \sigma_2$. These relations are: $[S_{12}(\hat r)]^2 = 6+2 \vec 
\sigma_1\cdot \vec \sigma_2-2 S_{12}(\hat r)$, $\{S_{12}(\hat r),\vec \sigma_1
\cdot \vec \sigma_2 \}=  2S_{12}(\hat r)$ and $ (\vec\sigma_1\cdot \vec\sigma_2
)^2=3-2\vec \sigma_1\cdot \vec \sigma_2$. 

\begin{figure}
\unitlength 1mm
\begin{picture}(160,55)
\put(0,0){\makebox{\psfig{file=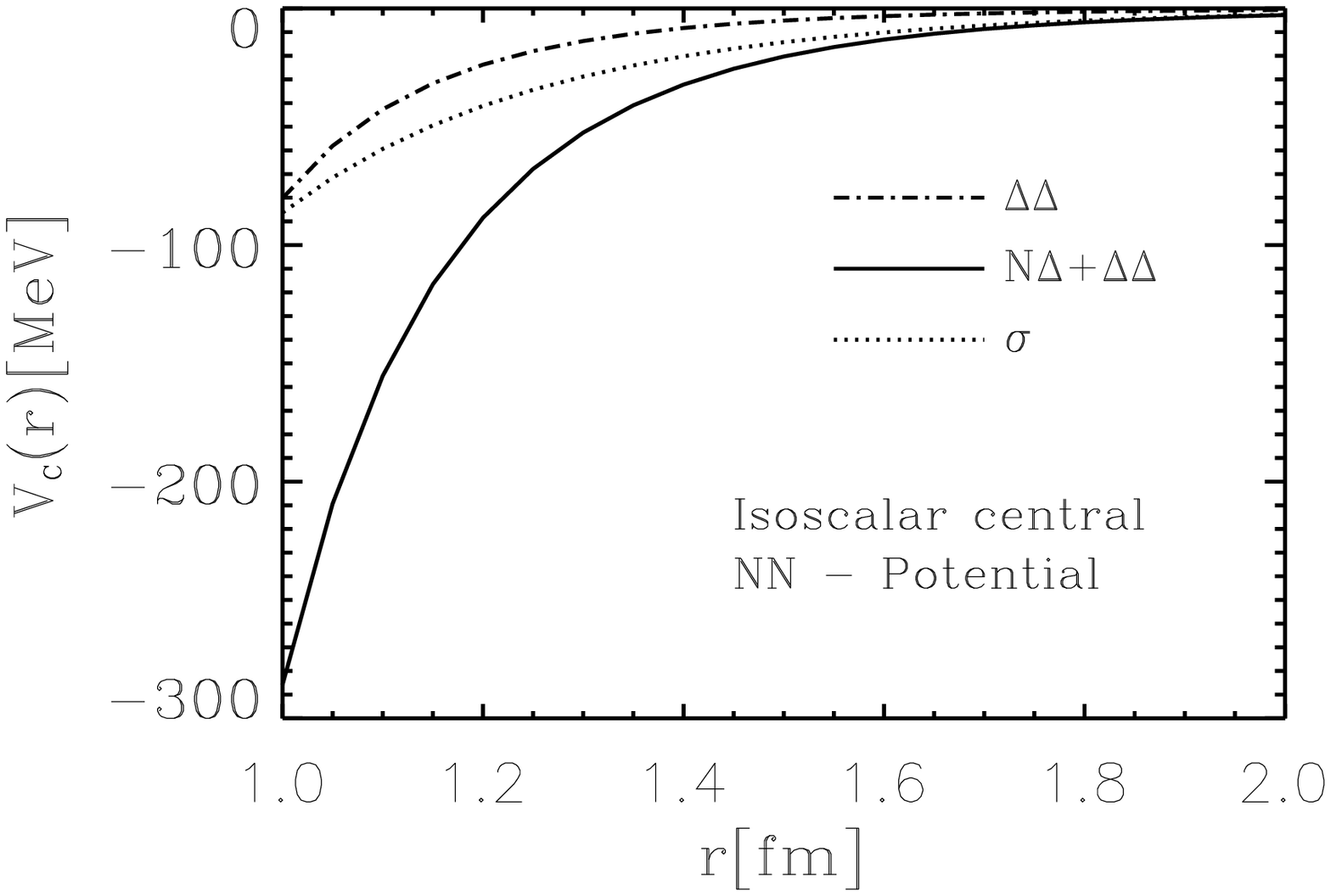,width=80.0mm}}}
\put(80,0){\makebox{\psfig{file=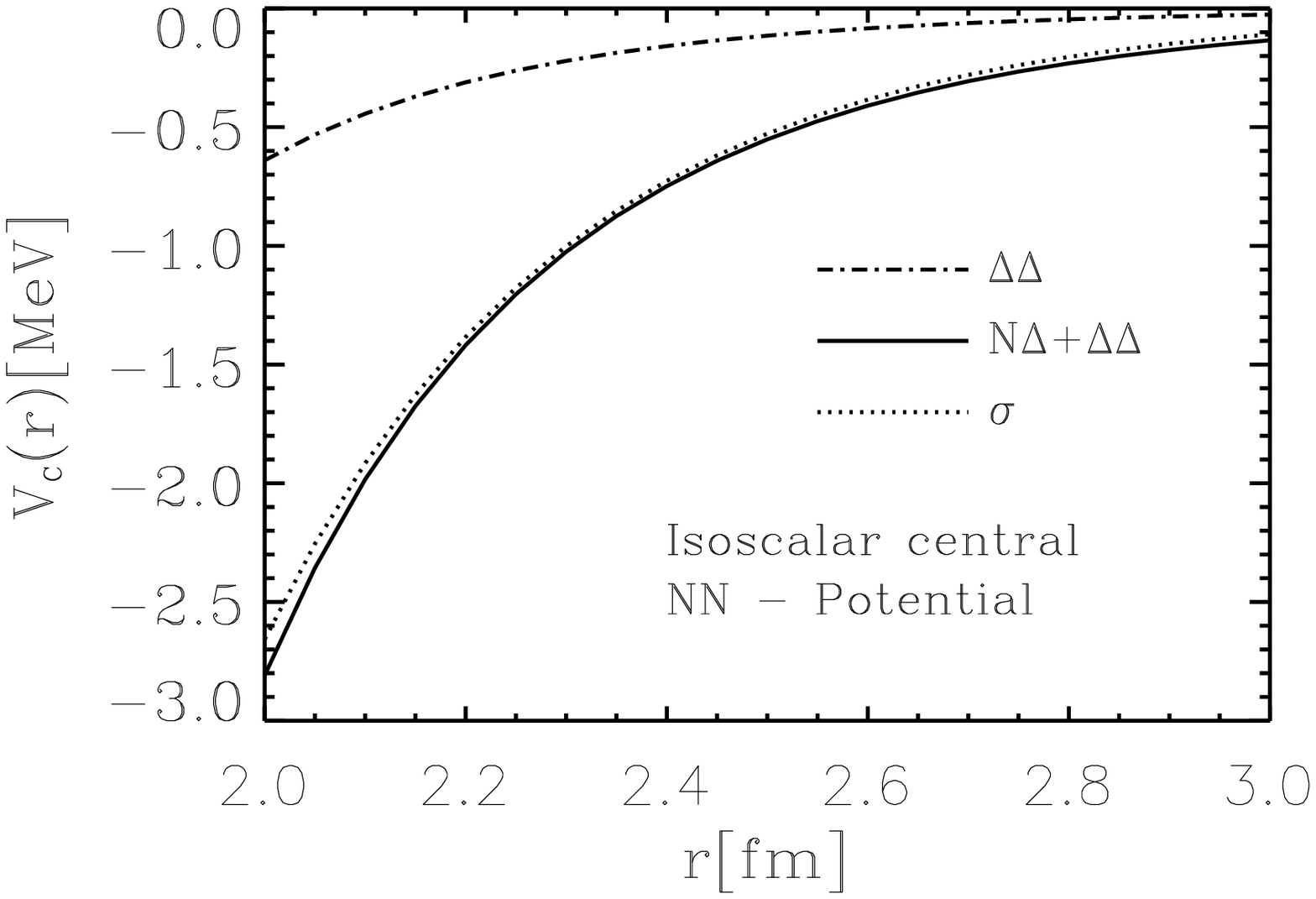,width=80.0mm}}}
\end{picture}
\end{figure}

\begin{center}
{\it Fig.2: The isoscalar central NN-potential $\widetilde V_C(r)$ generated 
by $2\pi$-exchange with $\Delta(1232)$-excitation versus the nucleon distance
$r$. }  \end{center}

\bigskip

Note that the relevant energy
difference $\Delta/12 = 24.4$ MeV in eq.(25) is rather small. It is quite 
remarkable that these simple relations between $\widetilde V_C^{(N\Delta)}(r),
\, \widetilde W_{S,T}^{(N\Delta)}(r)$ and  the $1\pi$-exchange potentials 
(eq.(25,26,27)) are exact for the isoscalar central and the isovector 
spin-spin/tensor potentials generated by the direct and crossed box graph with
single $\Delta$-excitation  (in the static limit $M\to \infty$). Heuristic 
considerations already suggest such a form of the isoscalar central 
$2\pi$-exchange potential.

The double $\Delta$-excitation diagrams amount to about $30\%$ of the total 
isoscalar central potential as shown by the dashed-dotted line in Fig.2. 
The fictitious "$\sigma$"-exchange potential $\widetilde V_C^{(\sigma)}(r) = 
-(g_\sigma^2/4\pi r) e^{-M_\sigma r}$ with $g_\sigma^2/4\pi= 7.1$ and
$M_\sigma=550$ MeV  \cite{bonn} is also shown  in Fig.2 by the dotted line. 
For distances $r>2$ fm one observes an almost perfect agreement between the
phenomenological "$\sigma$"-exchange potential and the total isoscalar central
potential generated by $2\pi$-exchange with $\Delta$-excitation. For shorter 
distances  $r<2$ fm the latter one increases more strongly in magnitude due to
its inherent $r^{-6}$-singularity (van der Waals behavior). Among all
potentials generated by $\Delta$-excitation the isoscalar central $\widetilde
V_C(r)$ is by far the  largest one. The attractive isovector central potential
$\widetilde W_C(r)$ is approximately a factor ten smaller, with values of
$-32.7$ MeV, $-2.0$ MeV and $-0.25$ MeV at distances $r= 1$ fm, $1.5$ fm and
$2.0$ fm.  The repulsive isoscalar and isovector spin-spin potentials
$\widetilde V_S(r), \, \widetilde W_S(r)$ and the attractive isoscalar and
isovector tensor potentials $\widetilde V_T(r),\, \widetilde W_T(r)$ are even
smaller, with typical values of $\pm 13$ MeV at $r=1$ fm. The asymptotic 
behavior for large $r$ is $e^{-2m_\pi r} r^{-5/2}$ for the isovector central
potential $\widetilde W_C(r)$  and $e^{-2m_\pi r} r^{-7/2}$ for the isoscalar
spin-spin/tensor potential $\widetilde V_{S,T}(r)$. In the other cases it can
be read off from eqs.(25,26,27). We note that throughout the large-$r$
asymptotics is determined alone by the box graphs with single
$\Delta$-excitation b). 

We have furthermore evaluated the first relativistic correction, proportional 
to $1/M$, to the $\Delta$-excitation graphs starting from the Rarita-Schwinger
form of the spin-3/2 propagator and $\pi N \Delta$-vertex. Explicit formulas 
for the respective NN-amplitudes can be found in the appendix. Unfortunately 
these corrections are not small in the range $1$ fm $< r<2$ fm or for $T_{lab}
> 100 $ MeV. The reason is probably found in combinatoric factors (up to six 
vertices and propagators are expanded in powers of $1/M$) and in the fact that
the ratio $\Delta/M \simeq 0.3$ is not so small. We expect that part of these 
relativistic $1/M$-corrections will be cancelled by higher orders in the 
$1/M$-expansion. On the other hand it is not clear how realistic a 
description of the $\Delta$-resonance via a local Rarita-Schwinger spinor-field
really is, even though it is the only available Lorentz-covariant formalism 
for spin-3/2 particles \cite{benmer}. Off its mass-shell the Rarita-Schwinger 
spinor-field exhibits potentially unphysical spin-1/2 degrees of freedom. In 
view of these problems we prefer not to include the relativistic 
$1/M$-correction to the $\Delta$-excitation graphs (Fig.1) in our calculation.

Let us give another argument in support of the idea that the 
$\Delta$-excitation in the static limit ($M\to \infty$) already represents the 
low energy $\pi \pi\to N\overline N$ dynamics in the scalar isoscalar channel
quite well. The shift of the nucleon scalar form factor $\sigma_N(t)$ from 
momentum transfer $t=0$ to $t=2m_\pi^2$ has been evaluated in 
ref.\cite{spectra} via dispersion relations and empirical $\pi N$- and 
$\pi\pi$-data.  The result is $\sigma_N(2m_\pi^2)-\sigma_N(0) = 15.2 \pm 0.4$ 
MeV. In the one-loop approximation of heavy baryon chiral perturbation theory
one finds for this  scalar-isoscalar quantity \cite{shift},
\begin{eqnarray} \sigma_N(2m_\pi^2)-\sigma_N(0) &=& {3g_A^2m_\pi^2\over 64\pi^2
f_\pi^2} \bigg\{ \pi \, m_\pi - 4 \sqrt{\Delta^2-m_\pi^2}
\ln{\Delta + \sqrt{\Delta^2-m_\pi^2} \over m_\pi} \nonumber \\ & & +(\pi -4) 
\Delta + 8 \Delta^3 \, D(i\sqrt2  m_\pi) \bigg\} = (8.0 + 6.9) \, {\rm MeV}
\end{eqnarray}
where the first term $\sim \pi m_\pi$ comes from the pion loop-diagram with a
nucleon intermediate state, and the remaining ones come from an analogous 
diagram with an intermediate $\Delta(1232)$-isobar ($8\Delta^2 D(i\sqrt2
m_\pi)=7.02$). The sum $14.9$ MeV of both terms agrees well with the 
dispersion-theoretical value $15.2 \pm0.4$ MeV. While the detailed spectral 
function Im\,$\sigma_N(t)$  given by these two one-loop diagrams is not in 
perfect agreement \cite{shift} with the empirical one derived via dispersion 
theory \cite{spectra}, the relevant integral over Im\,$\sigma_N(t)/t(t-2m_\pi^2
)$ is well reproduced.  

We expect a similar mechanism to be at work for the $2\pi$-exchange isoscalar 
central NN-potential which can anyhow only be tested at large and intermediate
distances, $r>1.5$ fm. The mass spectrum Im\,$V_C$ as given by 
the $\Delta$-excitation diagrams in the static limit does certainly not have 
all structures as the one derived from dispersion theory \cite{disp,hoehler}, 
in particular there is no enhancement around $550$ MeV (the broad
"$\sigma$"-meson). By analogy with the previous discussion of the nucleon 
scalar form factor we expect that the long and intermediate range components of
the isoscalar central potential $\widetilde V_C(r)$ are well represented by the
$\Delta$-excitation graphs in the static limit. For the long and intermediate 
range isoscalar central potential only the low energy part and some global 
features, but not the details of its mass spectrum, play a role. The good 
agreement between the "$\sigma$"-potential and the $\Delta$-excitation 
$2\pi$-exchange potential for $r>2$ fm in Fig.2  demonstrates this fact very 
clearly.

\section{CORRELATED TWO-PION EXCHANGE}

As previously mentioned the mass spectrum in the isoscalar central channel
Im\,$V_C\sim |f_{0+}|^2$, with the $\pi\pi\to N\overline N$ S-wave amplitude 
$f_{0+}$, shows an enhancement around $550$ MeV (the so-called
"$\sigma$"-meson) \cite{disp,hoehler}. This enhancement is often interpreted as
originating from correlated two-pion exchange (the first diagram shown in
Fig.3) in which the two exchanged  pions interact while propagating between the
two nucleons. This motivated us to evaluate this specific two-loop diagram in
heavy baryon chiral perturbation theory.

\begin{center}
\SetScale{0.8}
\SetWidth{1.5}
  \begin{picture}(364,72)
\Line(0,0)(0,85)
\Line(57,0)(57,85)
\DashLine(0,20)(57,66){6}
\DashLine(0,66)(57,22){6}
\Vertex(28.5,43){5}

\Line(100,0)(100,85)
\Line(157,0)(157,85)
\DashLine(100,42.5)(157,71){6}
\DashLine(100,42.5)(157,14){6}
\DashCArc(100,25)(17.5,90,270){6}

\Line(199,0)(199,85)
\Line(256,0)(256,85)
\DashLine(199,71)(256,42.5){6}
\DashLine(199,14)(256,42.5){6}
\DashCArc(256,42.5)(17.5,-90,90){6}

\Line(298,0)(298,85)
\Line(355,0)(355,85)
\DashCArc(312,42.5)(13,0,360){6}
\DashCArc(341,42.5)(13,0,360){6}
\Vertex(325.5,42.5){5}

\Line(397,0)(397,85)
\Line(454,0)(454,85)
\DashCArc(411,42.5)(13,0,360){6}
\DashLine(426,42.5)(454,66){6}
\DashLine(426,42.5)(454,20){6}
\Vertex(424.5,42.5){5}

  \end{picture}

\medskip

{\it Fig.3: Correlated two-pion exchange diagrams and some related two-loop
graphs. The dot represents the $\pi\pi$-interaction} 

\end{center}
  
\medskip

The correlated $2\pi$-exchange diagram involves the off-shell
$\pi\pi$-interaction which is, however, not uniquely defined by the effective
chiral Lagrangian. It depends on the choice of the interpolating pion field or,
to be explicit, on the parametrization of the SU(2)-matrix $U(\vec \pi\,)$  
which  is the basic variable entering the non-linear chiral
$\pi\pi$-Lagrangian, 
\begin{equation} U(\vec \pi\,) = 1+ {i\over f_\pi} \vec \tau \cdot \vec \pi -
{1\over 2 f_\pi^2} \vec \pi \,^2 - {i \, \alpha \over f_\pi^3} (\vec \tau \cdot
\vec \pi\,)^3 +{\alpha - {1\over 8} \over f_\pi^4} \vec \pi\,^4 + \dots
\end{equation} 
The arbitrariness of the coefficient $\alpha$ in the 3rd and 4th power of the
pion field just reflects this ambiguity. For the pure, isolated 
$\pi\pi$-interaction this does not cause a problem since the off-shell 
amplitude $A_{\pi\pi}= (s_{\pi\pi}-m_\pi^2 +2\alpha (4m_\pi^2-q_1^2-q_2^2-q_3^2
-q_4^2))/f_\pi^2$ is not an observable and the on-shell amplitude can be easily
shown to be independent of $\alpha$. However, the isoscalar central 
NN-amplitude $V_C$ generated by the first diagram in Fig.3 is 
$\alpha$-dependent. It is therefore not meaningful to consider this diagram 
alone, instead one has to find the complete subclass of diagrams 
for which the unphysical $\alpha$-dependence drops out. This  subclass is 
obtained by adding those diagrams which result from shifting the $4\pi$-vertex 
to a nucleon line (see a typical representative in Fig.3). This procedure 
resembles the construction of gauge invariant subclasses. The four additional 
diagrams now indeed cancel an unwanted term proportional to $10\alpha-1$. We 
note that for isovector amplitudes there is no $\alpha$-dependence in the 
first diagram of Fig.3. Altogether we find the following contribution to the 
NN T-matrix from the factorizable two-loop  diagrams in Fig.3:
\begin{eqnarray} V_C &=& -{3g_A^4 \over 1024\pi^2 f_\pi^6}(m_\pi^2+2q^2)
\Big[ m_\pi  + (2m_\pi^2+q^2) A(q)\Big]^2 \,\,, \\ 
W_C &=& -{1 \over 18432\pi^4 f_\pi^6}\bigg\{\Big[ 4m_\pi^2(1+2g_A^2)+q^2(1+5
g_A^2) \Big] L(q)\nonumber \\ &&-4m_\pi^2(1+2g_A^2)+ q^2(1+5g_A^2)\ln{m_\pi 
\over \lambda}-{q^2\over 6} (5+13g_A^2) \bigg\}^2 \,\,, \\
W_T&=&-{1\over q^2}\,W_S = -{g_A^4 \over 2048\pi^2 f_\pi^6}
\Big[ m_\pi+ w^2 A(q) \Big]^2 \,\,. \end{eqnarray}
Note that $V_C$ and $W_T$ are finite in dimensional regularization. The
divergent terms in $W_C$ have been omitted according to the usual minimal 
subtraction prescription which introduces a logarithmic dependence on the 
renormalization scale  $\lambda$. The expressions in eqs.(30,31,32) for 
correlated $2\pi$-exchange in heavy baryon chiral perturbation theory have a 
simple interpretation in terms of the one-loop contributions to the
nucleon scalar and isovector electromagnetic form factors \cite{kambor}:
\begin{eqnarray} V_C&=& {32\pi\over 3m_\pi^4}t_0^0(-q^2)\Big[\sigma_N(-q^2
)_{\rm loop}\Big]^2 \,\,, \\ W_C&=& -{1\over 8f_\pi^2} \Big[ G_E^V(-q^2)_{\rm
loop}  \Big]^2 \,\,, \\ W_T&=& -{1\over q^2}\, W_S=-{1\over 32 M^2 f_\pi^2}
\Big[ G_M^V(-q^2)_{\rm loop} \Big]^2 \,\,. \end{eqnarray}
with $t_0^0(s_{\pi\pi}) = (2s_{\pi\pi}-m_\pi^2)/(32\pi f_\pi^2)$ the 
on-mass-shell isospin-zero S-wave  $\pi\pi$-amplitude at leading
order. Actually, we have considered for $W_C$ a larger class of factorizable
two-loop graphs as shown in Fig.3, such that no additive constant to the
electric from factor $G_E^V(-q^2)_{\rm loop}$ appears under the square in
eq.(34). Note that both central NN-amplitudes in eqs.(33,34) are negative which
indicates a repulsive NN-interaction. The negative sign of $V_C$ comes from the
isospin-zero S-wave $\pi\pi$-amplitude evaluated at negative $s_{\pi\pi}=-q^2<
0$. As a matter of fact chiral soft pion theorems require the
$\pi\pi$-interaction to be of derivative-nature which unavoidably leads to
strong energy dependence of the isospin-zero S-wave amplitude. An immediate
consequence is that the attractive isospin-zero S-wave $\pi\pi$-interaction
above threshold ($s_{\pi\pi} \geq 4m_\pi^2$) switches  over to repulsion
sufficiently far below the threshold ($s_{\pi\pi}<0$). In the NN-scattering
process the exchanged pion pair belongs exclusively to the latter kinematical
region $s_{\pi\pi}<0$. This is the basic reason why we find here a (weakly)
repulsive isoscalar central potential from correlated $2\pi$-exchange.  

\bigskip

\renewcommand{\arraystretch}{1.3}

\begin{center}
\begin{tabular}{|c|c c c| c  c| }
    \hline
    \,\,$r$[fm] &  1.0 & 1.5 & 2.0 & $r\to 0$ & $r\to \infty $\\
    $\widetilde V_C$ &  0.731 & 0.090   & 0.016 & $r^{-4}$ & $\,e^{-2m_\pi r}
    r^{-2} \ln r\, $\\
    $\widetilde W_C$ &  3.811 & 0.571   & 0.092 & $\,r^{-7} \ln r\,$ &
    $e^{-2m_\pi     r} r^{-5/2}$ \\
    $\widetilde W_S$ &  --0.170 & --0.022   & --0.004 & $r^{-4}$ & $e^{-2m_\pi
    r} r^{-3}$\\
    $\widetilde W_T$ &  0.193 & 0.022   & 0.004 & $r^{-4}$ & $e^{-2m_\pi
    r} r^{-3}$\\
   \hline
  \end{tabular} 

\bigskip

{\it Tab.1: Coordinate space potentials from correlated $2\pi$-exchange in
units of MeV.}

\end{center}

\bigskip

In Tab.1 we display some values for the coordinate space potentials as they 
derive from our calculation of the correlated  $2\pi$-exchange using the 
corresponding imaginary parts (Im\,$[A(i\mu)]^2 = 2\,{\rm Re}\, A(i\mu)\, {\rm
Im}\, A(i\mu)$ etc.) to evaluate their spectral representations \cite{kbw}. In
order to get an estimate of the magnitude of $\widetilde W_C(r)$ we set
$\lambda= M_\omega=782$ MeV. Indeed both central potentials and the tensor
potential are repulsive but they are also negligibly small. We have confirmed
this furthermore by evaluating the contribution of the NN-amplitudes in
eqs.(30,31,32) to the  phase shifts and mixing angles. In the case of the
correlated $2\pi$-exchange isovector spin-spin potential one can even give an 
expression in closed form,    
\begin{eqnarray}\widetilde W_S(r) &=& {g_A^4 \over 6144\pi^3 f_\pi^6}{e^{-2x}
\over r^7} \Big\{ (15+30x+24x^2+8x^3)(\gamma_E+\ln4x) \nonumber \\ & &
+ (15-30x+24x^2-8x^3)e^{4x}E_1(4x)-4x(15+15x+8x^2+2x^3)\Big\}\,\,,
\end{eqnarray} 
with $x=m_\pi r$, $\gamma_E=0.5772...$ the Euler-Mascheroni number and $E_1(4x)
=\int^\infty_{4x}d\zeta \zeta^{-1} e^{-\zeta}$ the exponential integral 
function. The isovector spin-spin potential $\widetilde W_S(r)$ in eq.(36) has
only a  $r^{-4}$-singularity near the origin and it vanishes identically in the
chiral limit, $m_\pi=0$. The isoscalar central potential $\widetilde V_C(r)$ 
and the isovector tensor potential $\widetilde W_T(r)$ from correlated 
$2\pi$-exchange have actually the same features. In contrast to this the
isovector central potential $\widetilde W_C(r)$ from correlated $2\pi$-exchange
does not vanish in the chiral limit $m_\pi=0$ and it has a much stronger $r^{-7
}\ln r$-singularity near the origin. In Tab.1 the asymptotic behavior for large
$r$ is also given.  

In ref.\cite{bonn} further correlated $2\pi$-exchange diagrams with single and
double $\Delta$-excitation were considered. We restrict ourselves here to a
rough estimate of their effects. Obviously the contribution of the diagrams 
with $\Delta$-excitation is overestimated in the limit of zero mass splitting,
$\Delta= 0$. In that case only different spin/isospin factors show up in
comparison with the diagrams involving nucleon intermediate states. The 
isoscalar central amplitude $V_C$ in eq.(30) would simply be multiplied by a 
factor $9$. Comparing with the scalar form factor difference
$\sigma_N(2m_\pi^2)-\sigma_N(0)$ where the $\Delta$-isobar in the zero mass
splitting limit triples the pure nucleon-term, one expects an amplification 
factor around $4$ to be a more realistic estimate of $\Delta$-effects for 
finite mass splitting. In the case of the isovector tensor amplitude $W_T$ the
degenerate $\Delta$-isobars would  lead to a factor $9/4$ in eq.(32), and for
the isovector central amplitude $W_C$ the $g_A$-dependent terms in eq.(31)
would be cancelled by degenerate $\Delta$-isobars. Even including such 
amplification factors due to intermediate $\Delta$-isobars the correlated
$2\pi$-exchange is a negligibly small correction. This completes the discussion
of the correlated $2\pi$-exchange in heavy baryon chiral perturbation
theory. As a main result we find that this diagram cannot be identified with
the enhancement showing up in the empirical isoscalar central spectral function
$|f_{0+}|^2$. Analogous features have already been observed in the one-loop
calculation of the $\pi N$ scattering amplitude in ref.\cite{pin}. There the
(class of) diagrams with pion self-interaction reduces the spin- and isospin
averaged P-wave scattering volume $P_1^+$ as well as the nucleon axial
polarizability $\alpha_A$. A reduction of these quantities is just the opposite
of what one naively expects from scalar-isoscalar $\pi\pi$-correlations.

\section{VECTOR MESON EXCHANGE}
It is clear that the coupling of vector mesons ($\rho,\, \omega$) to nucleons 
has no direct relation to chiral symmetry. The vector meson mass scale actually
marks the kinematical endpoint of chiral dynamics with weakly interacting 
pions. However, vector mesons are an important ingredient for the understanding
of nucleon structure, as witnessed e.g. by the dispersion-theoretical analysis 
of the nucleon electromagnetic form factors \cite{hoehler,mmd} and the 
successful one-boson-exchange models of the NN-interaction
\cite{bonn}. In our study of the peripheral NN phases we find that the chiral
$2\pi$-exchange considered so far is insufficient to describe the empirical 
F-wave phase shifts above $T_{lab}=150$ MeV (see Fig.4 in our previous work 
\cite{kbw}). The source of the discrepancy is a much too strong attraction 
between the nucleons for distances $r<1.5$ fm. Naturally one expects that 
vector meson exchange produces the repulsive interaction at intermediate
distances which will be able to cure this problem.

The coupling of the $\rho$- and $\omega$-meson to the nucleon is characterized
by vector coupling constants $g_{\rho N,\omega N}$ and tensor-to-vector 
coupling ratios  $\kappa_{\rho,\omega}$. We will apply symmetry relations to 
minimize the number of free vector meson parameters. The KSFR relation $M_\rho
=M_\omega = \sqrt2 g_{\rho} f_\pi = 782$ MeV leads to $g_{\rho}=6.0$ for the 
universal  $\rho$-coupling constant. Via isospin considerations one obtains
$g_{\rho N} = g_{\rho}/2 = 3.0$ and SU(3) symmetry would predict $g_{\omega 
N}^{SU(3)} = 3 g_{\rho}/2$. However, all existing determinations of $g_{\omega 
N}$ point towards a larger value. We use here $g_{\omega N} = \sqrt 2g_{\omega 
N}^{SU(3)}= 12.7$ as obtained from a global fit of the NN F-waves. This
number is in the range of values found from forward NN-dispersion relations 
\cite{grein} ($g_{\omega N} = 10.1\pm 0.9$), the Bonn NN-potential \cite{bonn} 
($g_{\omega N} =11.5$) and the recent dispersion analysis of the nucleon 
electromagnetic form factors \cite{mmd} ($g_{\omega N} =20.9\pm 0.3$). For the 
tensor-to-vector coupling ratio we use the values of ref.\cite{bonn}, 
$\kappa_\rho=6$ and $\kappa_\omega =0$, which are also confirmed by the nucleon
electromagnetic form factor analysis of \cite{mmd} ($\kappa_\rho = 6.1
\pm 0.2,\,\kappa_\omega = -0.16\pm 0.02)$. The precise value of $\kappa_\omega 
$ is irrelevant for the NN F-wave phase shifts as long as it is sufficiently 
small, $|\kappa_\omega|<0.2$. The $\rho$-meson exchange between two nucleons 
leads to the following  isovector NN-amplitudes,
\begin{eqnarray} W_C & = & -{g_{\rho N}^2 \over 4M^2 (M_\rho^2 +q^2) } \bigg\{
\Big[2E -{q^2\over 2} \Big( {1\over E+M} +{\kappa_\rho \over M} \Big)\Big]^2
\nonumber \\ & & +(4p^2-q^2) \Big( {\kappa_\rho q^2 \over 4 M (E+M)} -1 \Big)^2
\bigg\} \,\,,\\
W_T &=&  -{1\over q^2} \, W_S =   -{g_{\rho N}^2(1+\kappa_\rho)^2 \over 4 M^2 
(M_\rho^2 +q^2)} \,\,, \\
W_{SO} & = & -{g_{\rho N}^2 \over 4M^2 (M_\rho^2 +q^2) } \bigg\{
\Big[2E -{q^2\over 2} \Big( {1\over E+M} +{\kappa_\rho \over M} \Big)\Big]
\Big( {1\over E+M} +{\kappa_\rho \over M} \Big) \nonumber \\ & & + \Big[ 1+ 
\kappa_\rho \Big( {E\over M} -{q^2 \over 4 M (E+M)}\Big) \Big] \Big( 2 -
{\kappa_\rho q^2 \over 2 M(E+M)} \Big) \bigg\} \,\,,\\
W_Q & = & -{g_{\rho N}^2 \over 4M^2 (M_\rho^2 +q^2) } \bigg\{{(4p^2-q^2) 
\kappa_\rho^2 \over 4M^2(E+M)^2} -\Big( {1\over E+M} +{\kappa_\rho \over M} 
\Big)^2 \nonumber \\ & &-  {2\kappa_\rho \over M(E+M)}\Big[ 1+ \kappa_\rho 
\Big( {E\over M} -{q^2 \over 4 M(E+M)}\Big)  \Big] \bigg\} \,\,,
\end{eqnarray}    
where we have kept the fully relativistic expressions ($E=\sqrt{M^2+p^2}$).  
In a non-relativistic truncation at order $M^{-2}$ one would  otherwise loose
important contributions, in particular the quadratic spin-orbit term $W_Q$
proportional to the large coefficient $2\kappa_\rho^2+2 \kappa_\rho+1/4$. 
Interestingly, the fully relativistic tensor and spin-spin terms $W_{T,S}$ in 
eq.(38) agree exactly  with the lowest order non-relativistic approximation. 
For a pseudoscalar meson exchange this is also the case \cite{kbw}.  The
isoscalar NN-amplitudes due to $\omega$-exchange are obtained in complete
analogy by the replacement ($g_{\rho N}, \kappa_\rho) \to (g_{\omega N},
\kappa_\omega)$ in eqs.(37-40). For $\omega$-exchange (with $\kappa_\omega
\simeq 0$) the truncation at order $M^{-2}$ is sufficiently accurate. At
energies where $\rho$- and $\omega$-exchange are important it is not meaningful
to approximate them by local (polynomial) contact terms, since the ratio
$q/M_\omega$ is not small.  

For the sake of completeness one should also add the exchange of $\eta$-mesons
with mass $m_\eta = 547.45$ MeV.  In the absence of a reliable empirical
determination of the $\eta N$-coupling constant $g_{\eta N}$ we will use the
SU(3)-value  $g_{\eta N}=(3F-D)M/\sqrt 3 f_\pi = 4.4$ together with the
approximate values of the octet axial vector coupling constants $D=3/4$ and $
F=1/2$. For comparison the Bonn OBE-model (without $2\pi$-exchange) uses 
$g_{\eta N}=6.8$. The $\eta$-exchange leads to an isoscalar tensor amplitude of
the form, 
\begin{equation}V_T= {3 \over 64f_\pi^2 (m_\eta^2+q^2)}\,\,.\end{equation}
The actual calculation of the phase shifts shows that $\eta$-exchange
with the coupling strength given by SU(3) is almost negligible in all partial
waves with $L\geq 2$. We have furthermore evaluated the $K\overline K$-exchange
(bubble and triangle) diagrams in heavy baryon chiral perturbation theory. As
expected these processes lead to very small repulsive isoscalar and isovector
central NN-potentials, $\widetilde V_C(1\,{\rm fm}) = 0.99$ MeV, $\widetilde
W_C(1\, {\rm fm}) =0.26$ MeV.

\section{RESULTS FOR PHASE SHIFTS AND MIXING ANGLES}
In this section we present and discuss our results for the NN phase-shifts with
$L\geq 3$ and mixing angles with $J\geq 3$ up to nucleon laboratory kinetic
energies of $T_{lab} = 350$ MeV. For the D-wave phase shifts and $\epsilon_2$
we show results only up to $T_{lab}=120$ MeV. First, we state all 
ingredients which go into the calculation. We include all terms derived in the
recent work \cite{kbw}, the point-like $1\pi$-exchange, the iterated
$1\pi$-exchange and the irreducible $2\pi$-exchange with the low energy 
constants $c_{1,3,4}$ set equal to zero. The terms proportional to $c_{3,4}$ 
are now substituted by explicit $\Delta(1232)$-dynamics, and $c_1$ gave anyhow
only a marginal contribution to the isoscalar central amplitude $V_C$. The new
ingredients are  the $2\pi$-exchange with $\Delta$-excitation in the static
limit (section 3)  and  the vector meson exchange (section 5) with coupling
constants $g_{\rho N}=3, \,\kappa_\rho = 6,\, g_{\omega N}=12.7,\,
\kappa_\omega=0$. We use throughout the value $m_\pi=138$ MeV for the (average)
pion mass.

\subsection{D-WAVES} 

The D-wave phase shifts and mixing angle $\epsilon_2$ are shown in Fig.4 up to
$T_{lab}= 120$ MeV. The dashed curve corresponds to the one-pion exchange
approximation and the full curve includes in addition two-pion exchange and
vector meson exchange. The dotted curve represents the recent empirical
energy dependent NN phase shift analysis of ref.\cite{arnd} (VPI). The
triangles and squares give NN phase shifts and mixing angles derived
from single energy analyses of ref.\cite{dubois} and ref.\cite{arnd}, 
respectively. The circles represent the results of the multi-energy partial
wave analysis of ref.\cite{swart} (Tab.IV,V). In all cases the two-pion and 
vector meson exchange corrections go into the right direction, but deviations
show up already above $T_{lab}=30$ MeV in the $^1D_2$ partial wave and above
$T_{lab}= 50$ MeV for $^3D_1$ and $^3D_3$. The
$^3D_2$ phase shift and the mixing angle $\epsilon_2$ are in agreement with the
data up to $T_{lab}=100$ MeV. Similar results were found recently in
ref.\cite{ballot} using a somewhat different approach to the $2\pi$-exchange.
Compared to our previous calculation \cite{kbw}, there is no improvement in 
the D-wave phase shifts and $\epsilon_2$ due to adding (perturbative) $\rho$- 
and $\omega$-exchange. The $2\pi$-exchange with its $r^{-6}$ singular behavior
still provides too large attraction at distances $r\leq 1$ fm. Obviously, the 
D-wave phase shifts and $\epsilon_2$ above $T_{lab}=100$ MeV are already 
sensitive to the short range NN-repulsion beyond $\omega$-exchange. It appears,
that the D-waves (above $T_{lab}=100$ MeV), as well as the S- and P-waves, 
require non-perturbative methods and phenomenological parametrizations of the
short range NN-interaction. This is, of course, well known from earlier
investigations. 

\subsection{F-WAVES}   
The F-wave phase shifts and the mixing angle $\epsilon_3$ are shown in Fig.5.
We present here results up to $T_{lab}=350$ MeV, i.e. $70$ MeV above the
$NN\pi$-threshold where inelasticities are still negligible. 
The phase shifts in the $^1F_3, \, ^3F_2,\, ^3F_3$ partial waves are in very 
good agreement with the empirical data up to $T_{lab}=350$ MeV, whereas
deviations show up in the $^3F_4$ partial wave above $T_{lab}=220$ MeV. The
good agreement in the former case results from the inclusion of $\rho$- and
$\omega$-exchange. The $\omega$-exchange compensates the too strong attraction
from to $2\pi$-exchange. The $\rho$-exchange with its large tensor, spin-orbit and quadratic spin-orbit amplitudes ($\kappa_\rho=6$) leads to the
correct splitting of the singlet and the three triplet F-wave phase shifts. In
particular, the correct downward bending of the $^3F_2$ phase shift is a vector
meson exchange effect (see Fig.4 in \cite{kbw} where the opposite behavior was
found from $2\pi$-exchange alone). The deviation in the $^3F_4$ partial wave
above $T_{lab}=220$ MeV suggests that further short range effects are at work
in this particular channel. The mixing angle  $\epsilon_3$ is in perfect 
agreement with  the data for all energies up to $T_{lab}=350$ MeV. Irreducible 
$2\pi$-exchange and iterated $1\pi$-exchange contribute to this quantity
with roughly equal strength. There is also a small but non-negligible 
contribution from  vector meson exchange to $\epsilon_3$ above $T_{lab}=200$
MeV.  

\subsection{G-WAVES}   
The G-wave phase shifts and the mixing angle $\epsilon_4$ are shown in Fig.6. 
The predictions are in good agreement with the data for all four partial
wave phase shifts, and with the mixing angle up to $T_{lab}=350$ MeV. The 
vector mesons $\rho$ and $\omega$ almost do not operate anymore in these high
angular momentum states, $L=4$. At $T_{lab}=300$ MeV they produce phase shift
contributions of $0.3^\circ$ and smaller.  One has now reached the chiral 
window in which  $1\pi$- and (chiral) $2\pi$-exchange describe the 
NN-interaction completely and reliably. Note that the differences between 
$1\pi$-exchange and data are still sizeable in the $^1G_4$ and $^3G_5$ partial 
waves. The chiral $2\pi$-exchange closes this gap between the data and the 
$1\pi$-exchange approximation.  In the $^1G_4$ partial wave the correction 
comes mainly from irreducible $2\pi$-exchange, whereas in the $^3G_5$ partial 
wave (with total isospin $I=0$) both irreducible $2\pi$-exchange and iterated
$1\pi$-exchange contribute with roughly equal strength. Compared to our 
previous calculation \cite{kbw} the (very small) $^3G_5$ phase shift does not 
bend over to positive values any more. For the $^3G_3$
and $^3G_4$ phase shifts and $\epsilon_4$ the $2\pi$-exchange corrections are
relatively small. Nevertheless, these small effects improve the agreement
between data and chiral predictions.

\subsection{H-WAVES}
The H-wave phase shifts and the mixing angle $\epsilon_5$ are shown in Fig.7. 
The $^1H_5$ phase shift and the mixing angle $\epsilon_5$ have converged to the
$1\pi$-exchange approximation. In the $^3H_4$ and $^3H_5$ partial wave one 
finds small corrections to $1\pi$-exchange which nevertheless improve the 
agreement with the empirical data. Note however that the $1\pi$-exchange
approximation considerably underestimates the empirical $^3H_6$ phase shifts. 
Again this gap is closed by the chiral $2\pi$-exchange. It is rather remarkable
that the $2\pi$-exchange effects are still important at such a large orbital 
angular momentum, $L=5$. 

\subsection{I-WAVES}
The I-wave phase shifts and the mixing angle $\epsilon_6$ are shown in Fig.8.
Again, the $^3I_6$ phase shift and the mixing angle $\epsilon_6$ have converged
to the $1\pi$-exchange. In the $^1I_6$, $^3I_5$  and $^3I_7$ partial waves we
predict small differences to the empirical phases of ref.\cite{arnd}. The
calculation presented here is indeed most reliable in the high angular momentum
partial waves. 

\subsection{INTERACTION DENSITIES}
In order to learn about the relevant length scales at which the peripheral NN 
interaction actually takes place it is most instructive to study the local
interaction densities. These are  expressions of the form $r^2 j_L(pr)^2\,
\widetilde U_{LSJ}(r)$, with  $j_L(pr)$ a spherical Bessel
function and $\widetilde U_{LSJ}(r)$ the coordinate space potential in a given
state $|LSJ\rangle$. Unfortunately, a fully equivalent representation of the 
phase shifts $\delta_{LSJ}$ and the mixing angles $\epsilon_J$ in terms of 
local interaction densities is not possible because of some inherent 
non-localities of the iterated $1\pi$-exchange (section 4.3 in ref.\cite{kbw})
and the quadratic spin-orbit interaction. Nevertheless, the generic features
are already displayed by the dominant contributions to the NN T-matrix, namely 
$1\pi$-exchange and the isoscalar central components of $2\pi$- and 
$\omega$-exchange. Two examples of such interaction densities are  shown in
Fig.9 for the $^3F_4$ partial wave at $T_{lab}=200$ MeV and the $^1G_4$ partial
wave at $T_{lab}=250$ MeV. One can clearly see the peaking of the interaction
densities around distances of about $r=1.5$ fm and $r=2.5$ fm, respectively. 
It results on one side from the centrifugal barrier effect given by the wave 
function $r^2 j_L(pr)^2$ and on the other side from the exponential decay of 
the potential $\widetilde U_{LSJ}(r)$. Varying the orbital angular momentum $L$
and the center of mass momentum $p$, the peak of the interaction density moves
and appears approximately at a distance $r \approx L/p$. This length scale 
corresponds just to the classical impact parameter.

\section{CONCLUDING REMARKS}
The chiral perturbation theory calculation of the NN phase shifts and mixing
angles,  presented here, is most reliable in the high angular momentum partial
waves which probe the long and medium  range parts of the NN force. This is the
region where $1\pi$- and $2\pi$-exchange explains the NN-interaction in a 
model independent way. We have demonstrated that the S- and P-wave chiral
dynamics of the pion-nucleon system determines the peripheral NN phase shifts
almost completely, with no arbitrary cutoffs or "form factors". The new aspect
emphasized in this work is that the chiral pion-baryon effective Lagrangian
provides a well-defined systematic framework to deal with  the peripheral
nucleon-nucleon interaction. In particular there is no need to introduce the
scalar-isoscalar "$\sigma$"-meson. The intermediate range isoscalar central
attraction is explicitly produced by van der Waals-type $2\pi$-exchange
including intermediate $\Delta$-isobar excitations. Effects from
$\pi\pi$-rescattering turn out to be negligibly small, in accordance  with the
suppression of higher loops in chiral perturbation theory.

\begin{figure}
\unitlength 1mm
\begin{picture}(160,165)
\put(0,110){\makebox{\psfig{file=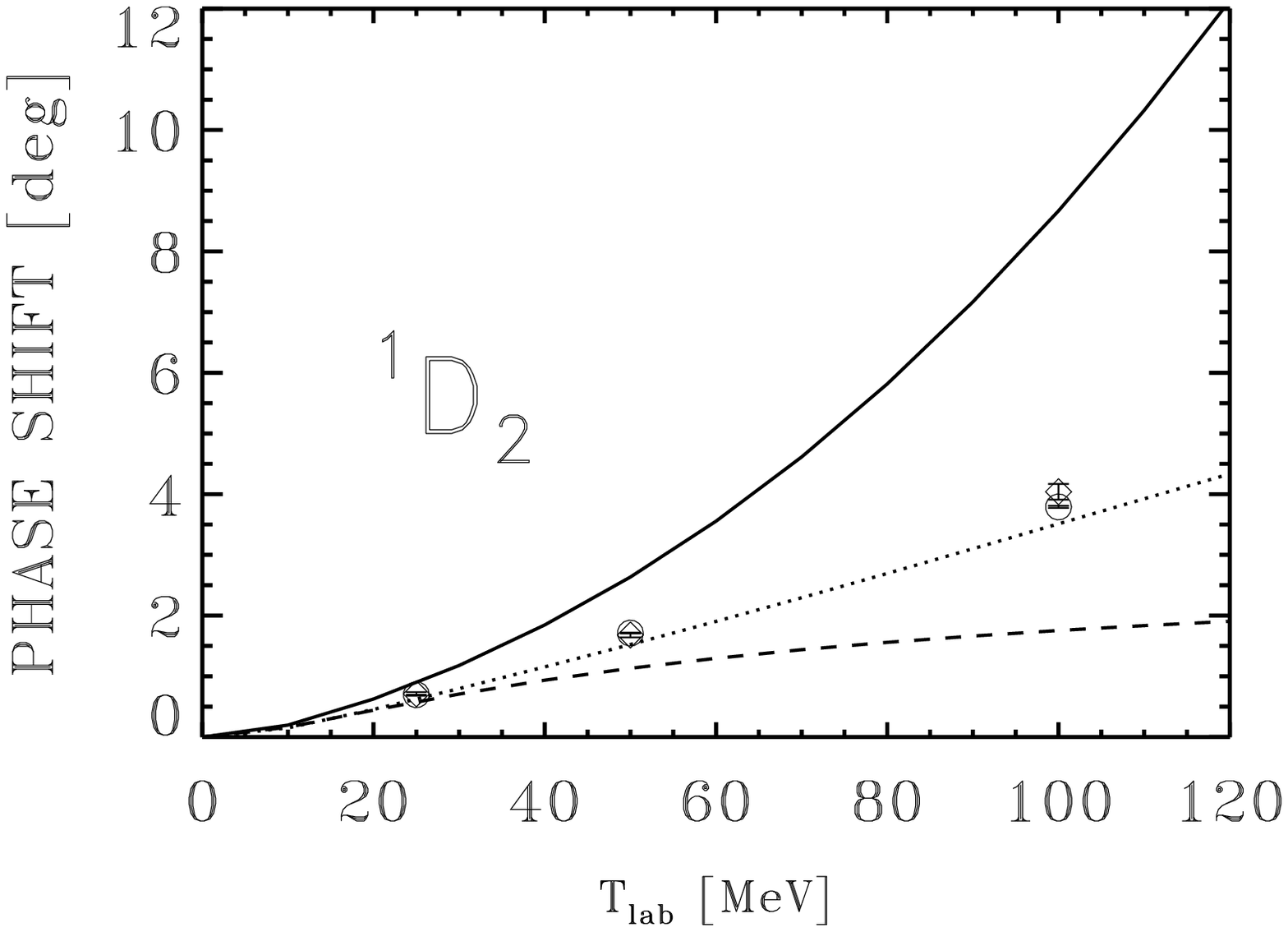,width=80.0mm}}}
\put(80,110){\makebox{\psfig{file=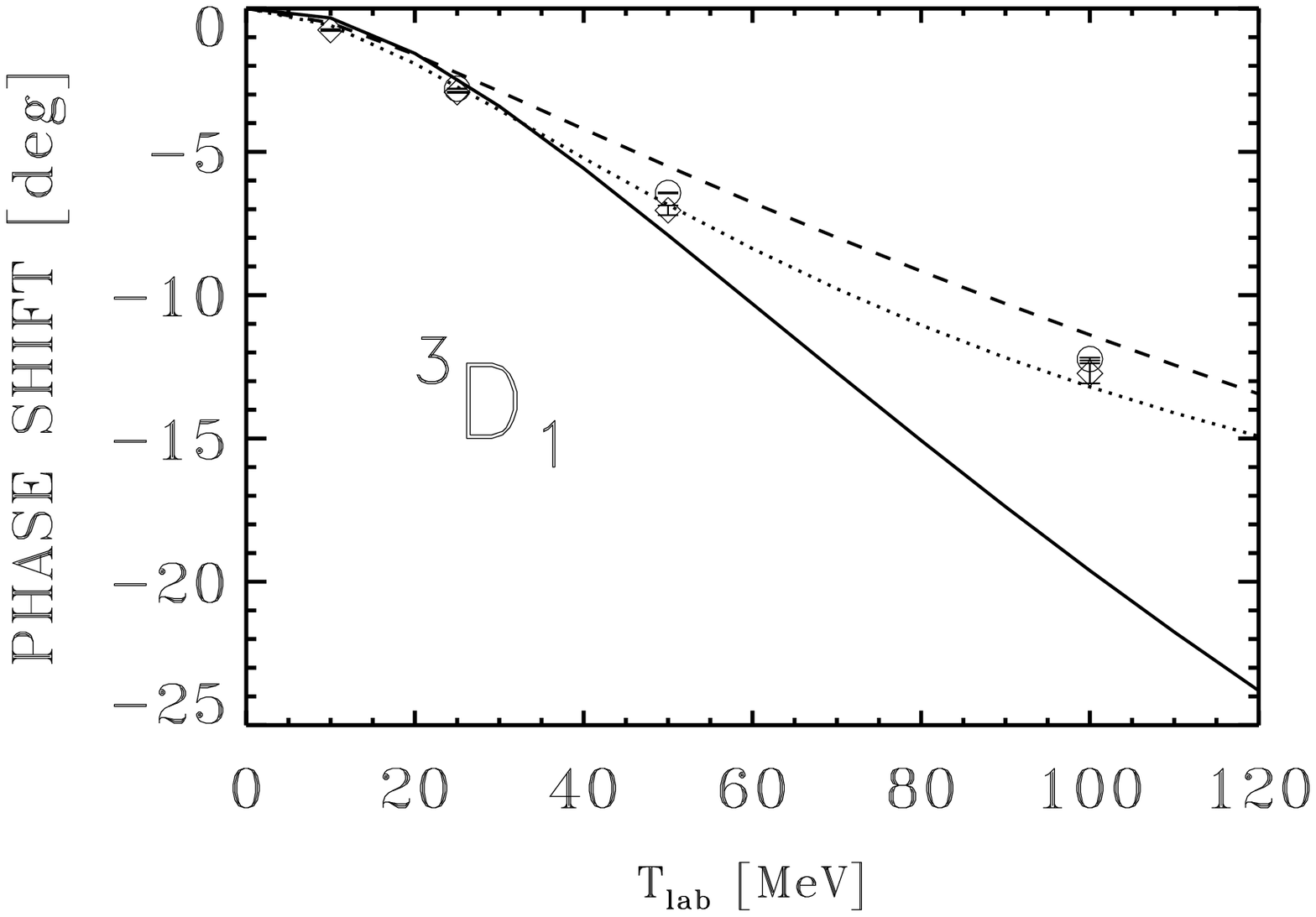,width=80.0mm}}}
\put(0,55){\makebox{\psfig{file=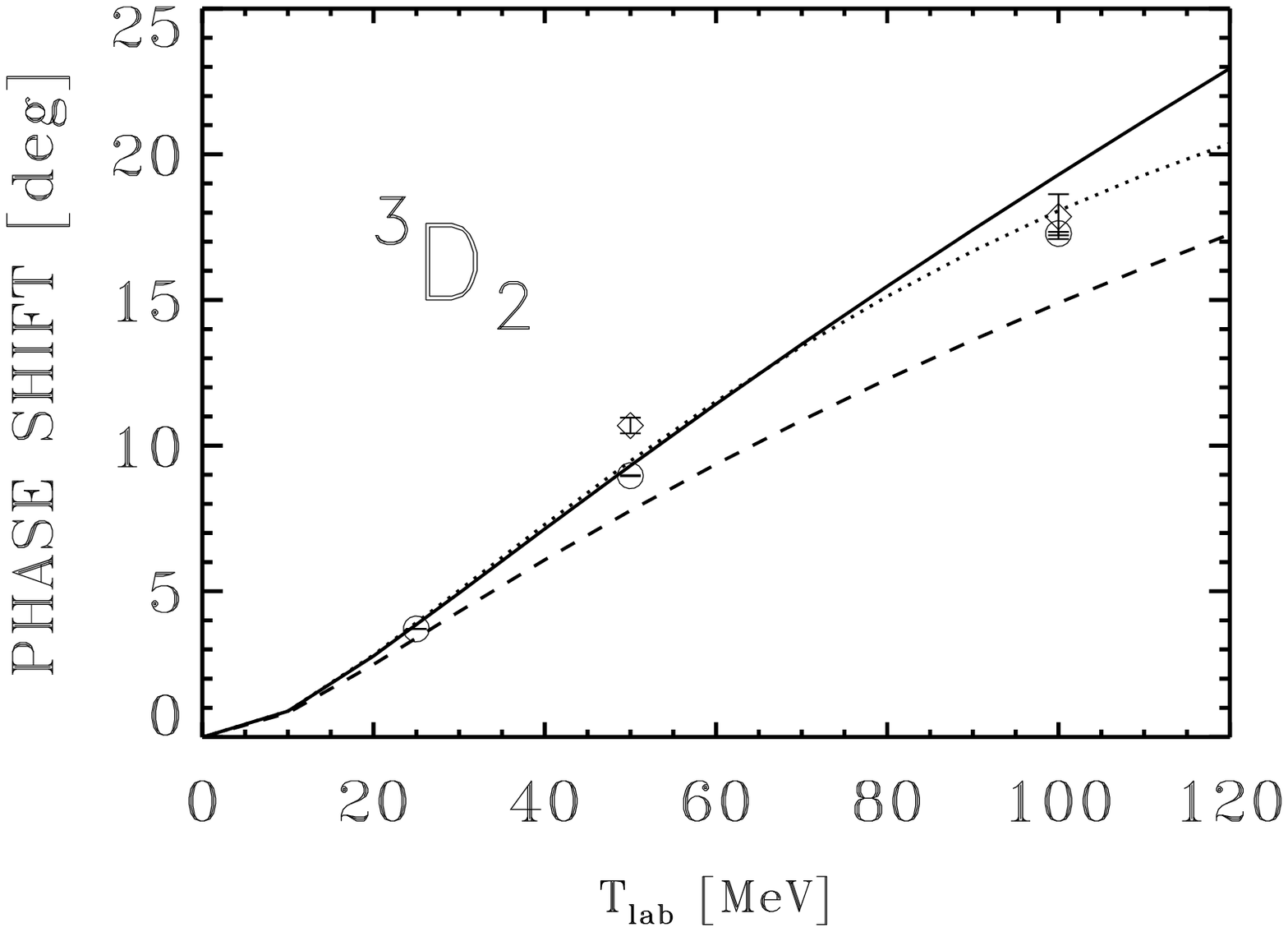,width=80.0mm}}}
\put(80,55){\makebox{\psfig{file=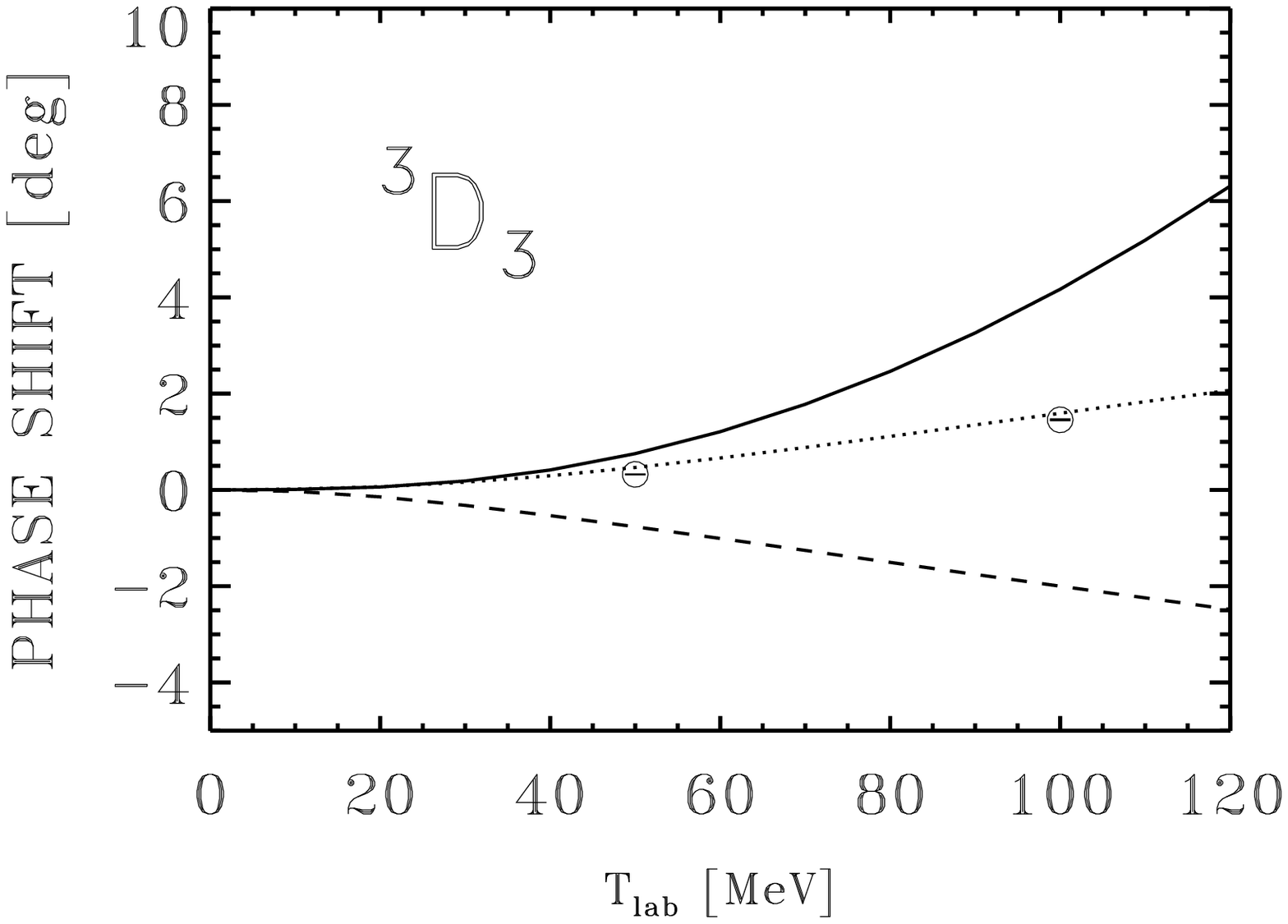,width=80.0mm}}}
\put(40,0){\makebox{\psfig{file=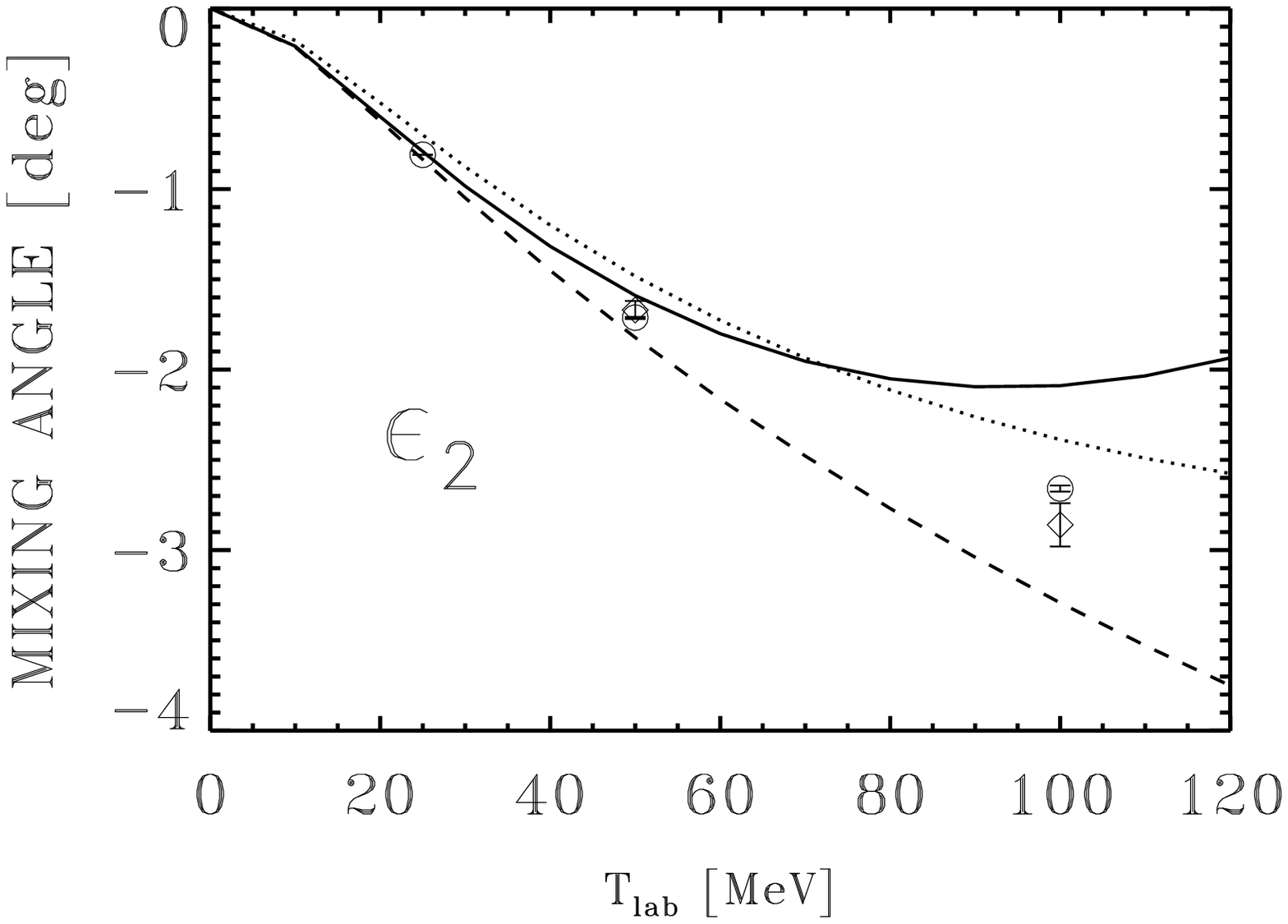,width=80.0mm}}}
\end{picture}
\end{figure}

{\it Fig.4: D-wave  NN phase shifts and mixing angle $\epsilon_2$ versus 
the nucleon laboratory kinetic energy $T_{lab}$. The dashed curves correspond
to the $1\pi$-exchange approximation and the full curves include chiral
$2\pi$- and vector meson exchange as well. The dotted curves represent the
empirical energy dependent NN phase shift analysis of ref.\cite{arnd}.}

\begin{figure}
\unitlength 1mm
\begin{picture}(160,165)
\put(0,110){\makebox{\psfig{file=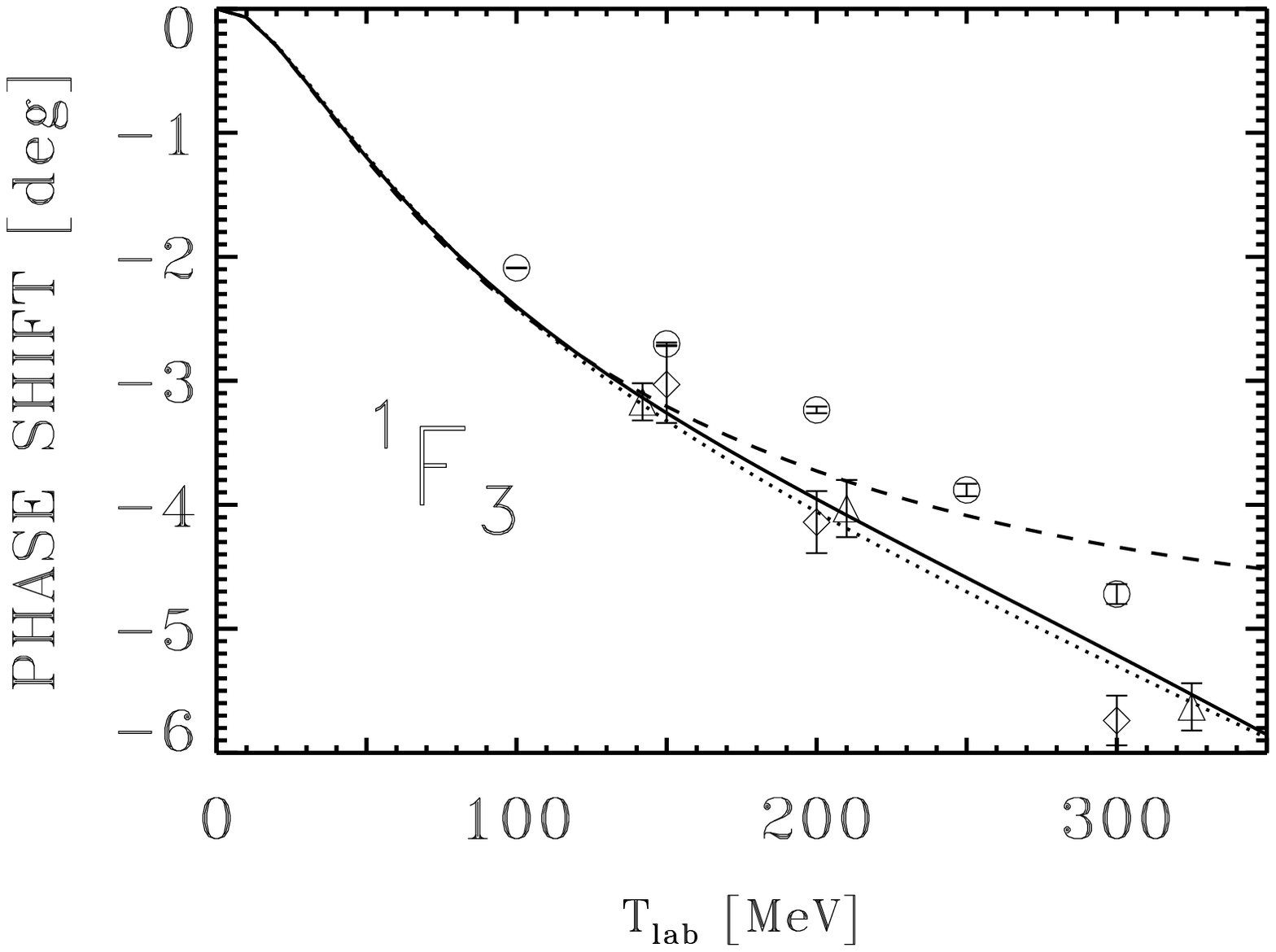,width=80.0mm}}}
\put(80,110){\makebox{\psfig{file=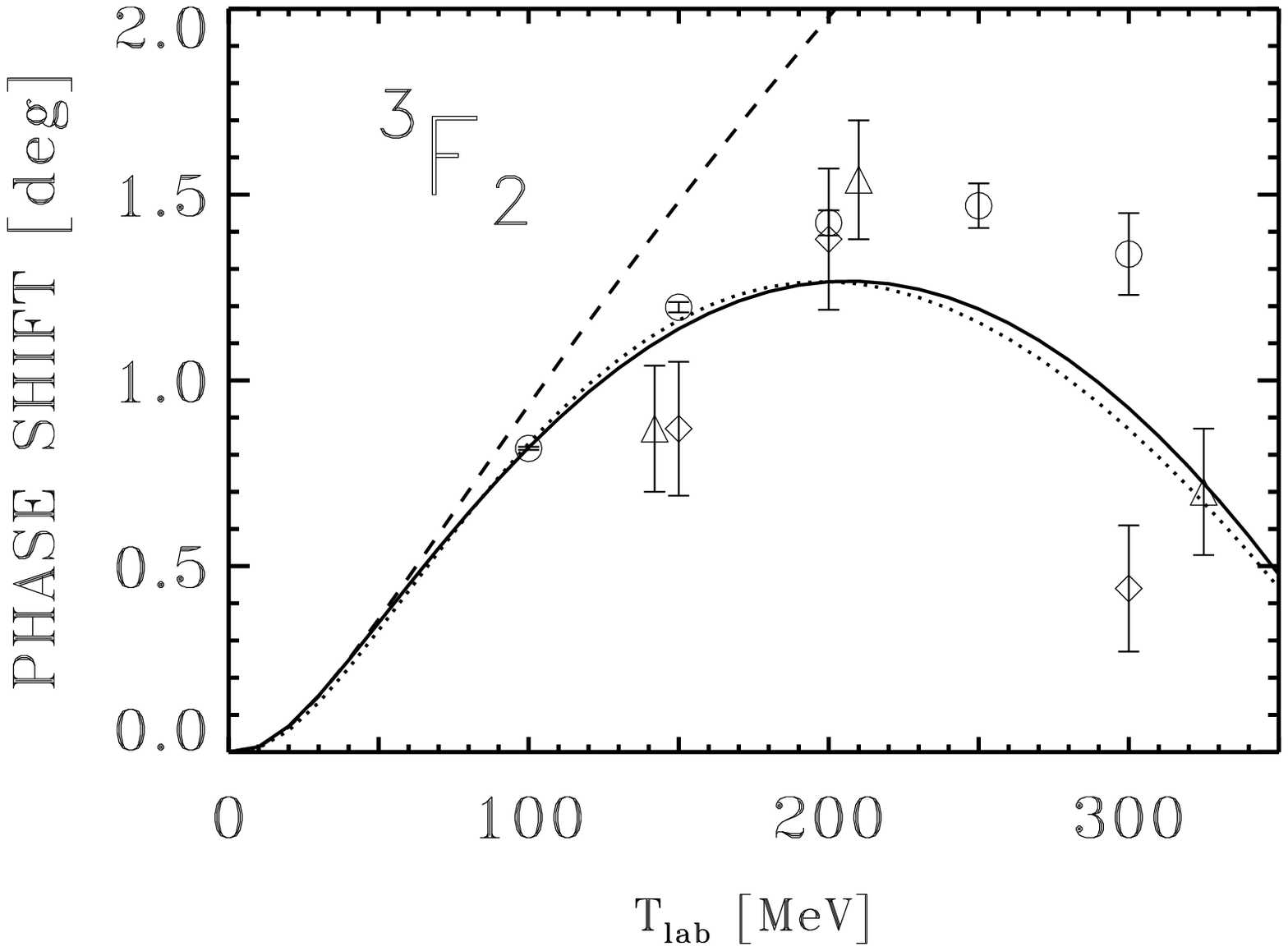,width=80.0mm}}}
\put(0,55){\makebox{\psfig{file=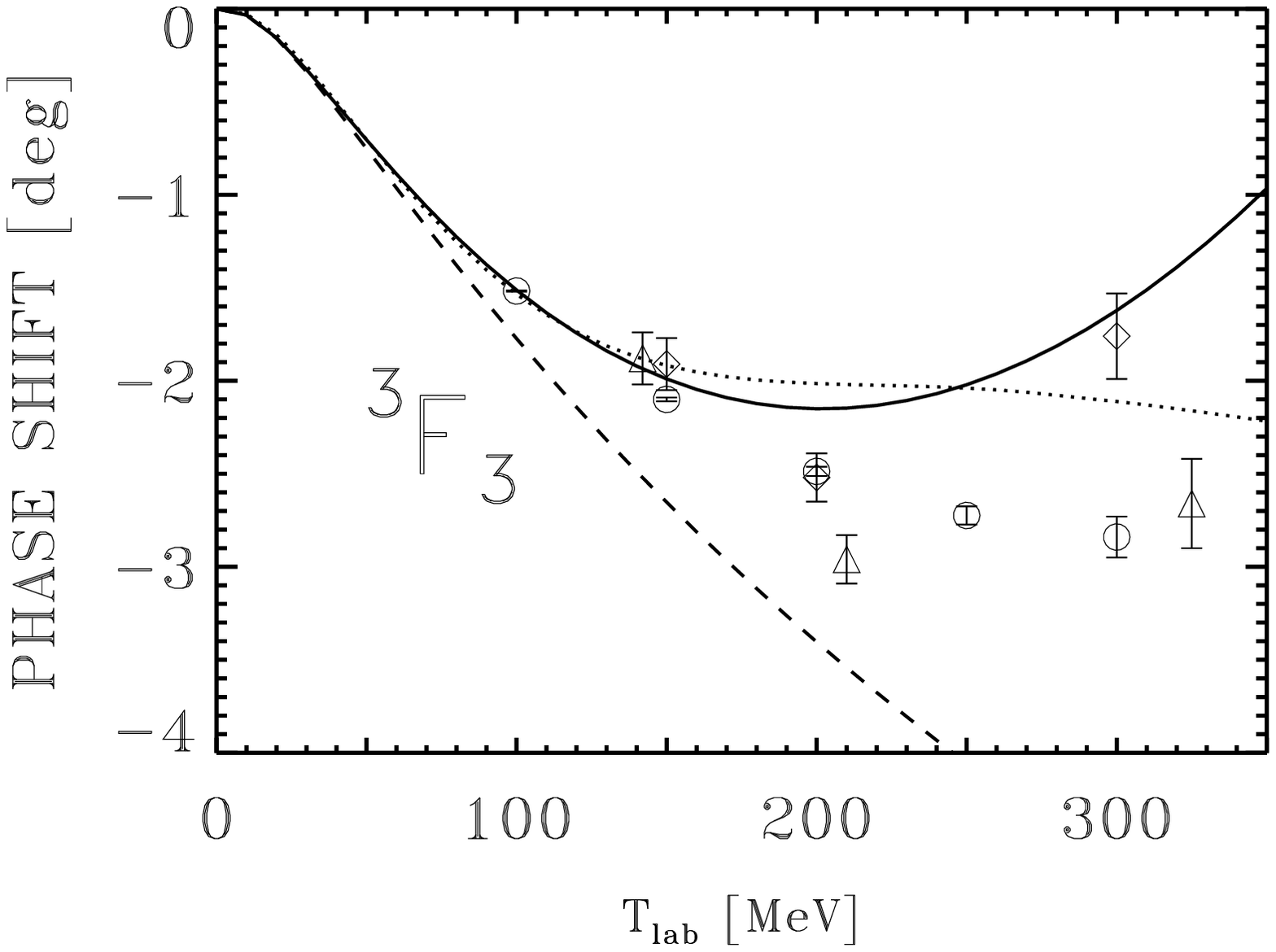,width=80.0mm}}}
\put(80,55){\makebox{\psfig{file=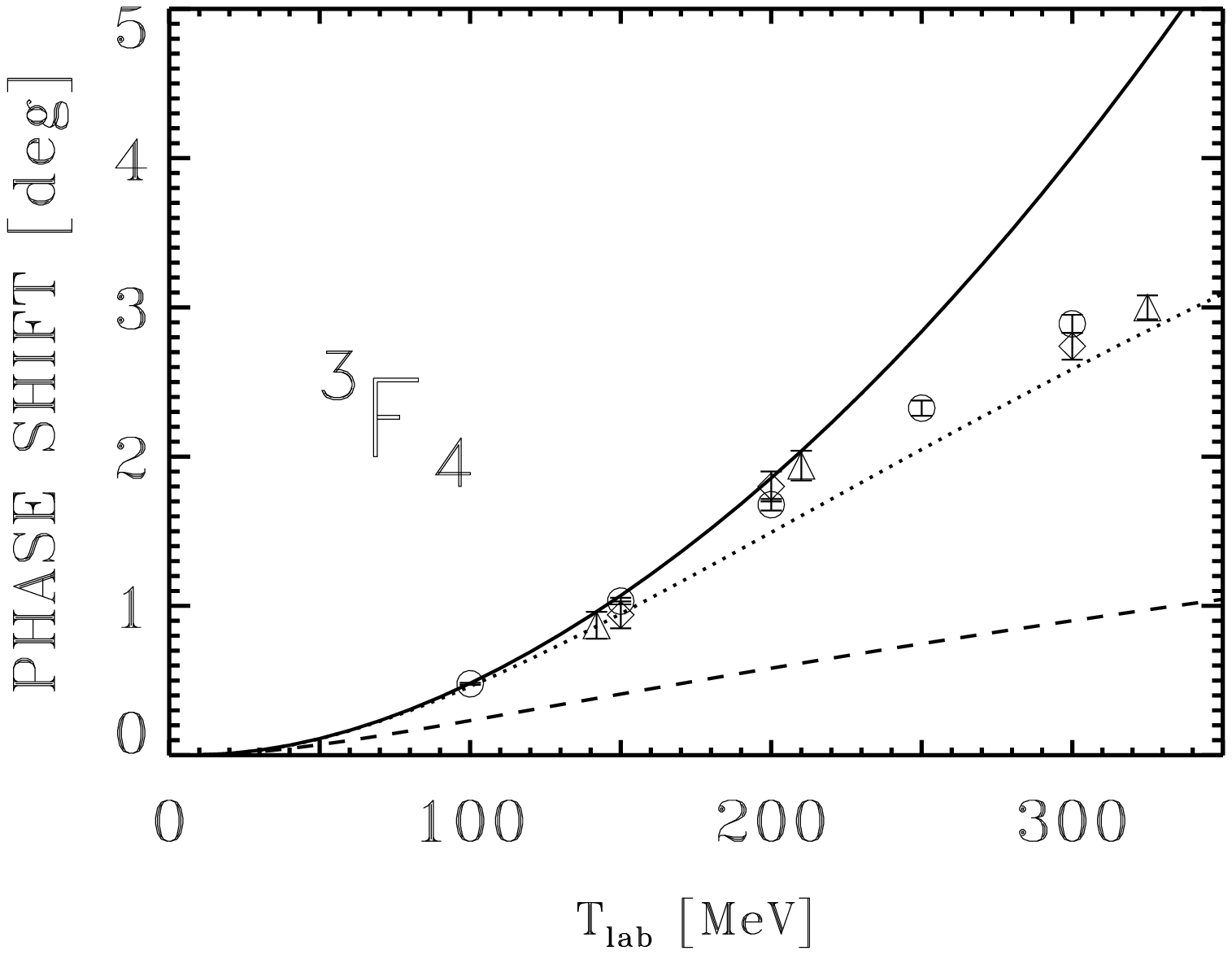,width=80.0mm}}}
\put(40,0){\makebox{\psfig{file=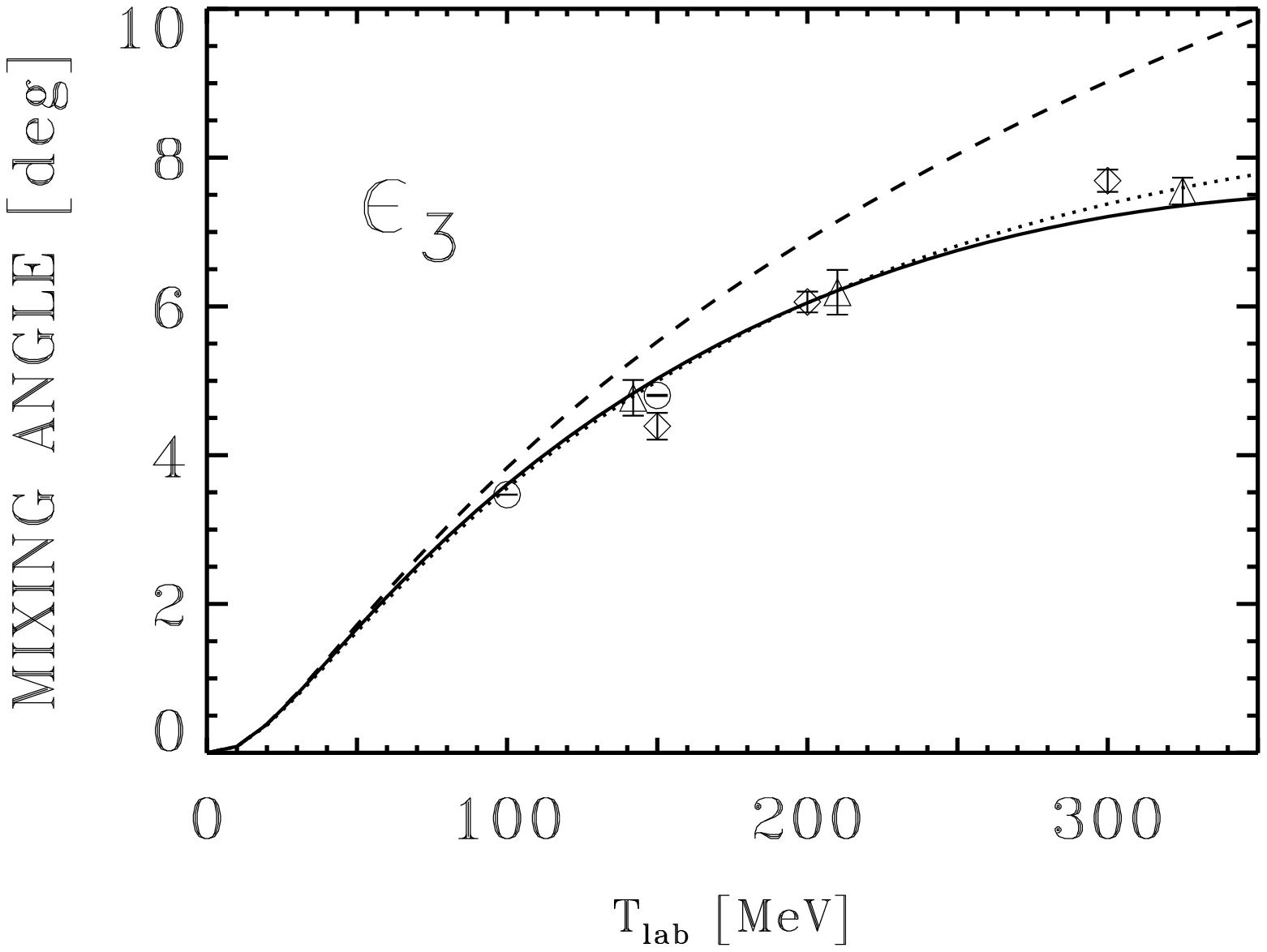,width=80.0mm}}}
\end{picture}
\end{figure}

{\it Fig.5: F-wave  NN phase shifts and mixing angle $\epsilon_3$ versus the
nucleon laboratory kinetic energy $T_{lab}$. For notations see Fig.4.}

\begin{figure}
\unitlength 1mm
\begin{picture}(160,165)
\put(0,110){\makebox{\psfig{file=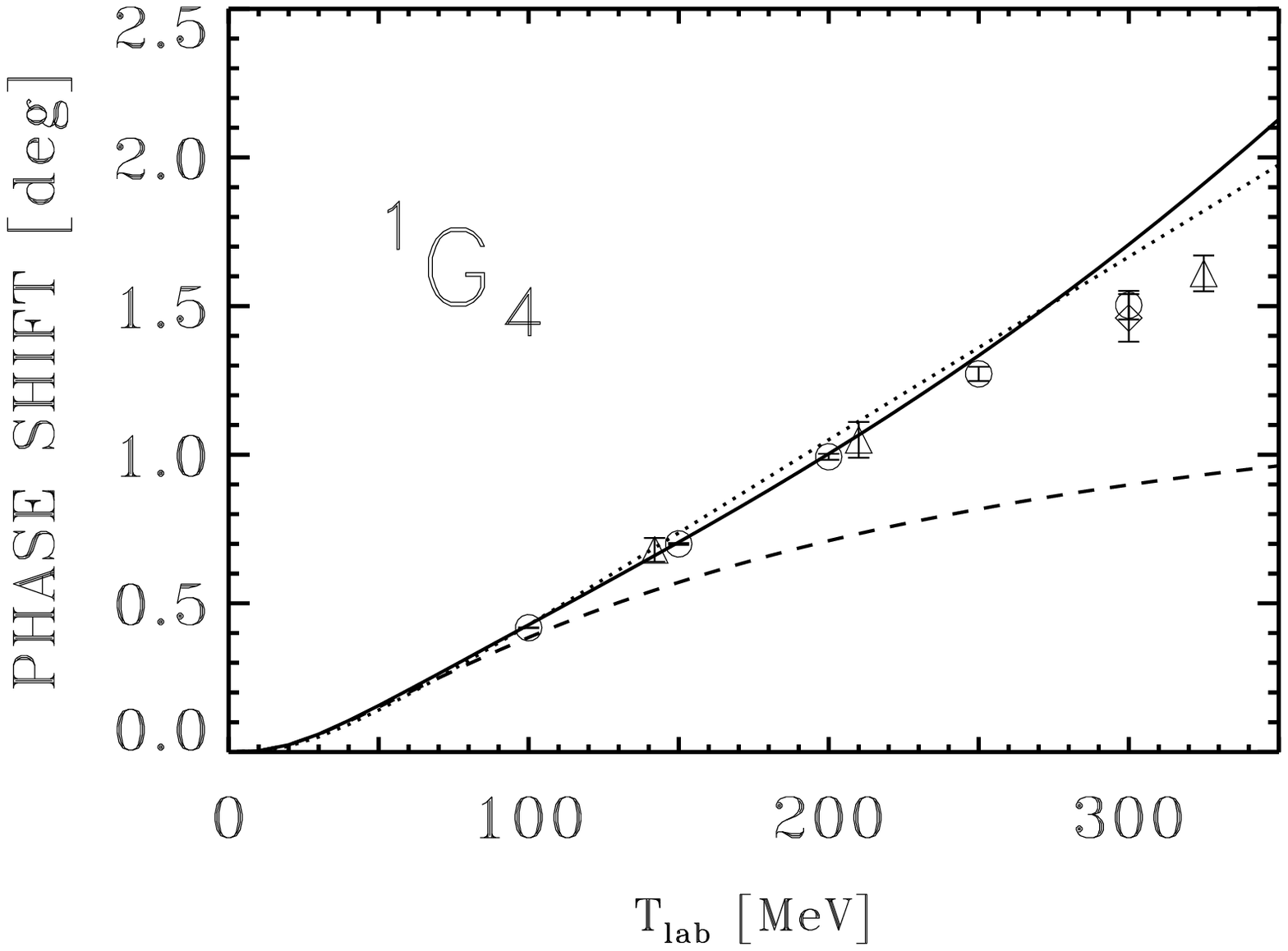,width=80.0mm}}}
\put(80,110){\makebox{\psfig{file=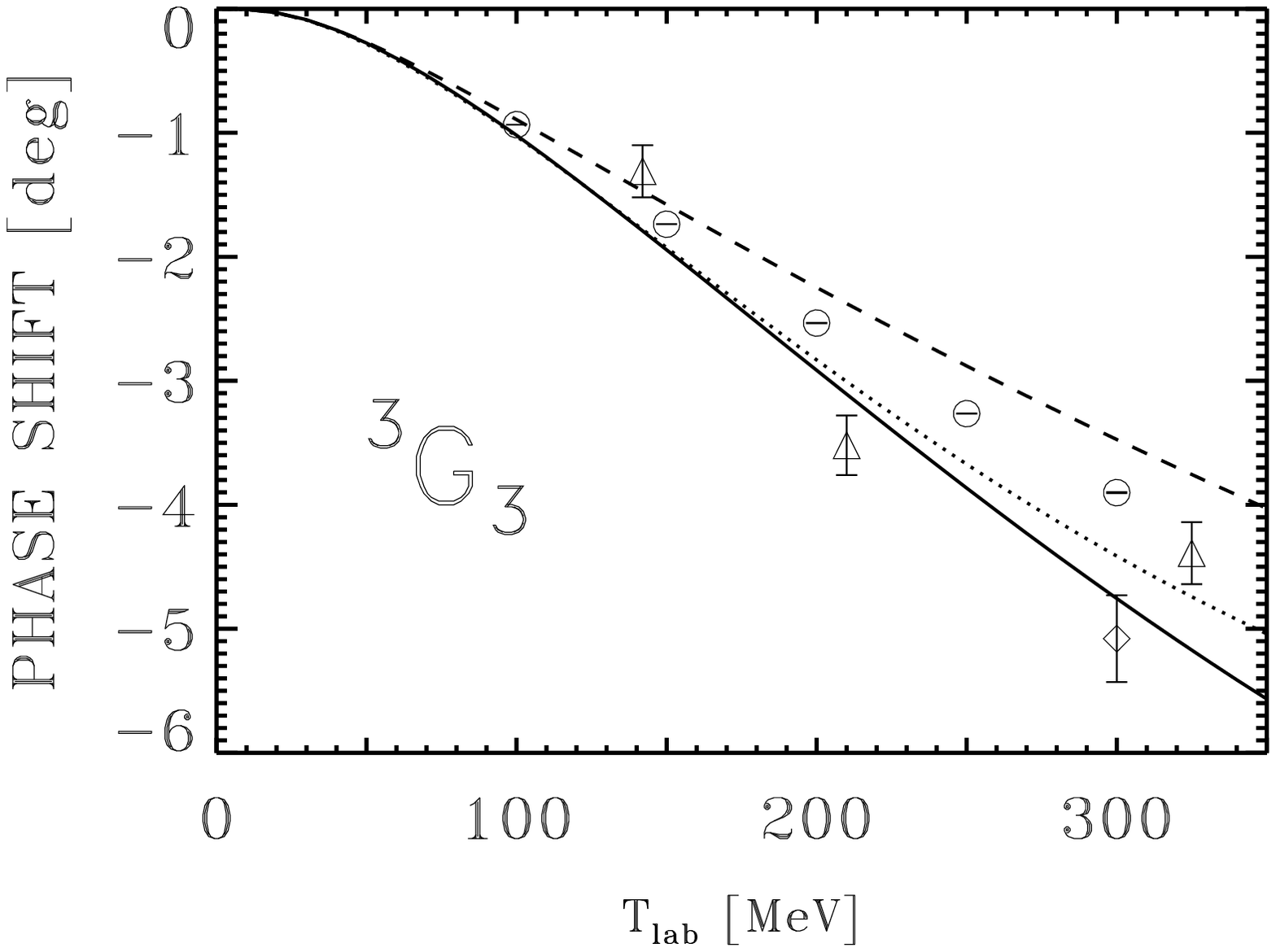,width=80.0mm}}}
\put(0,55){\makebox{\psfig{file=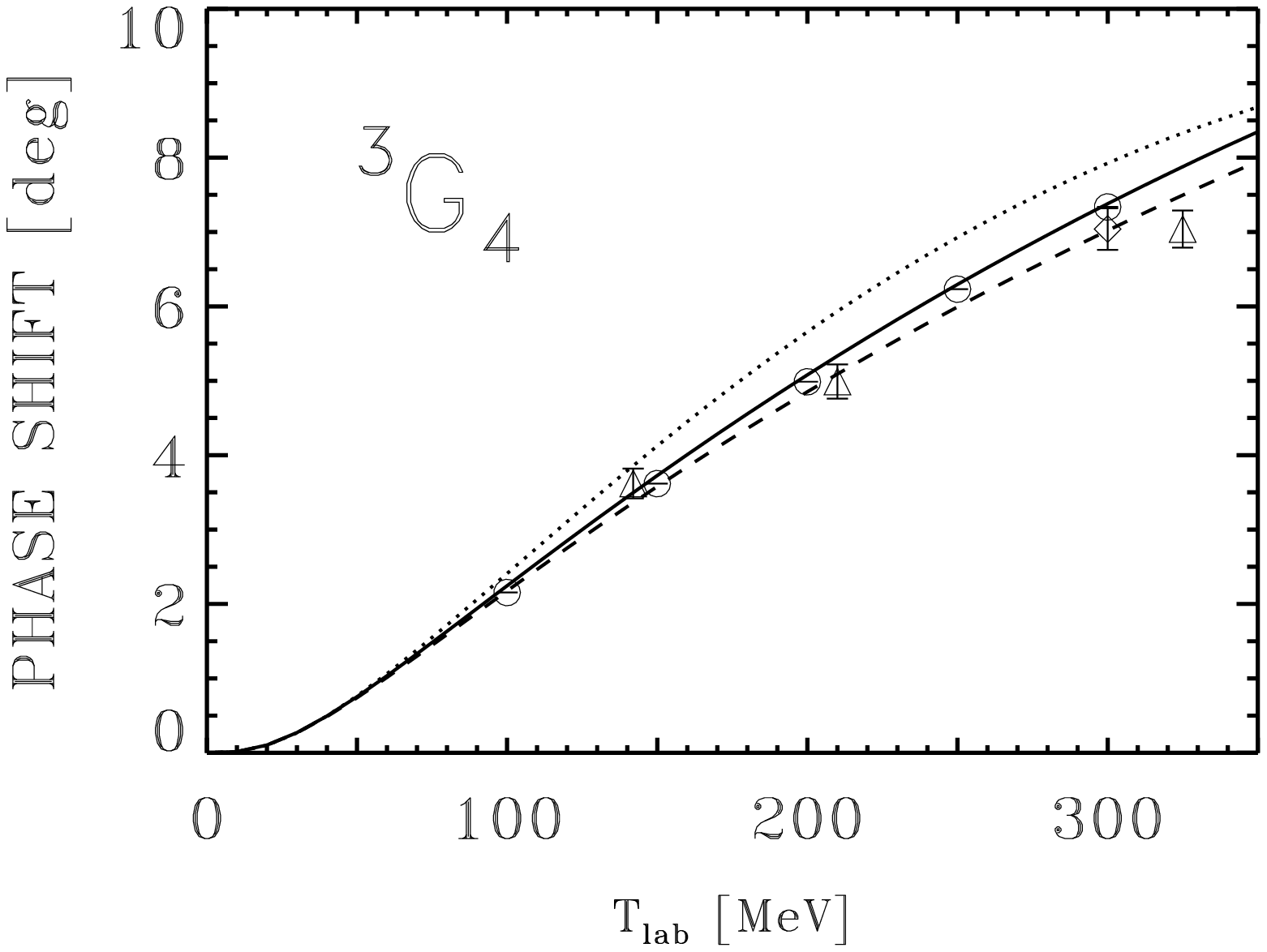,width=80.0mm}}}
\put(80,55){\makebox{\psfig{file=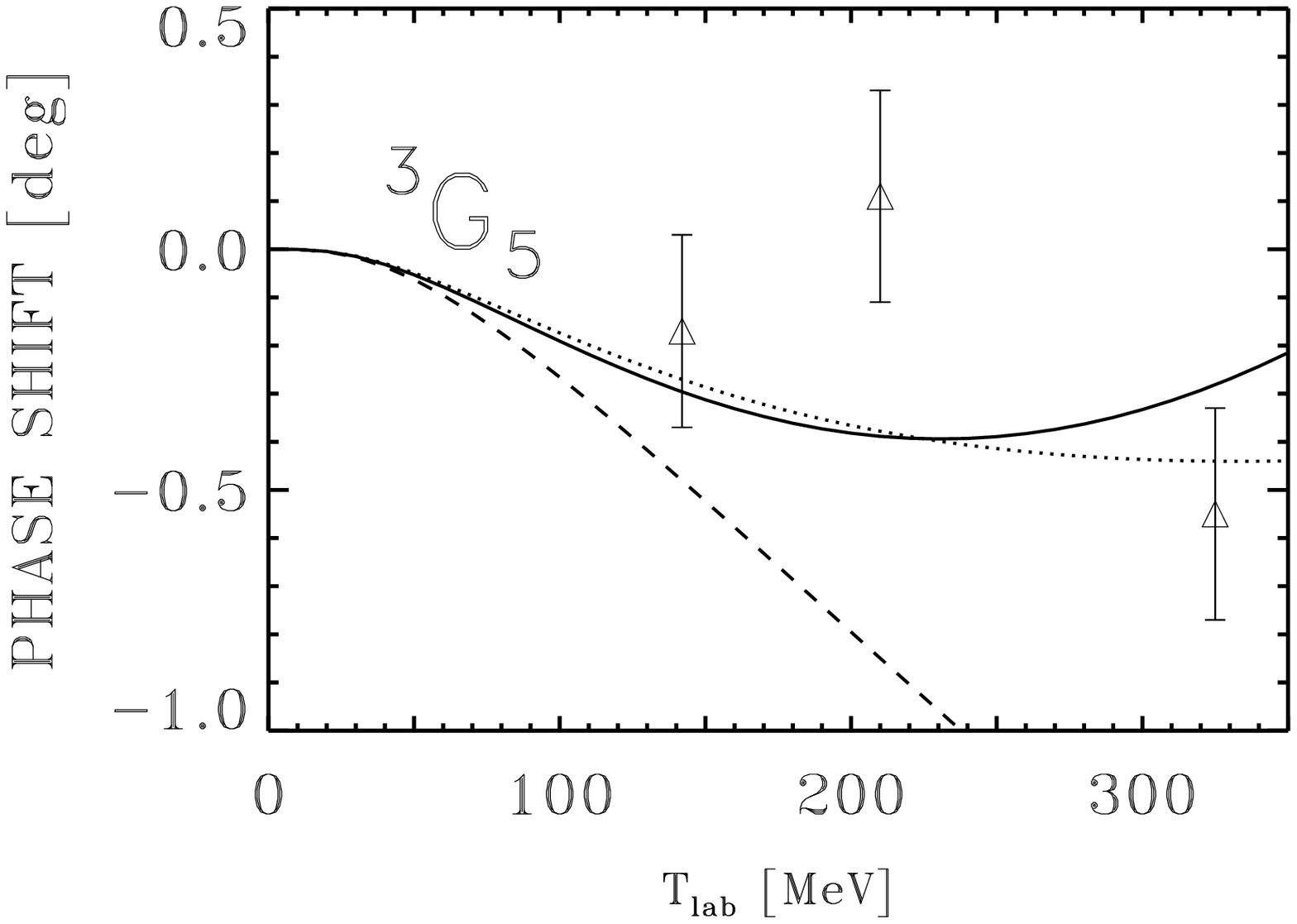,width=80.0mm}}}
\put(40,0){\makebox{\psfig{file=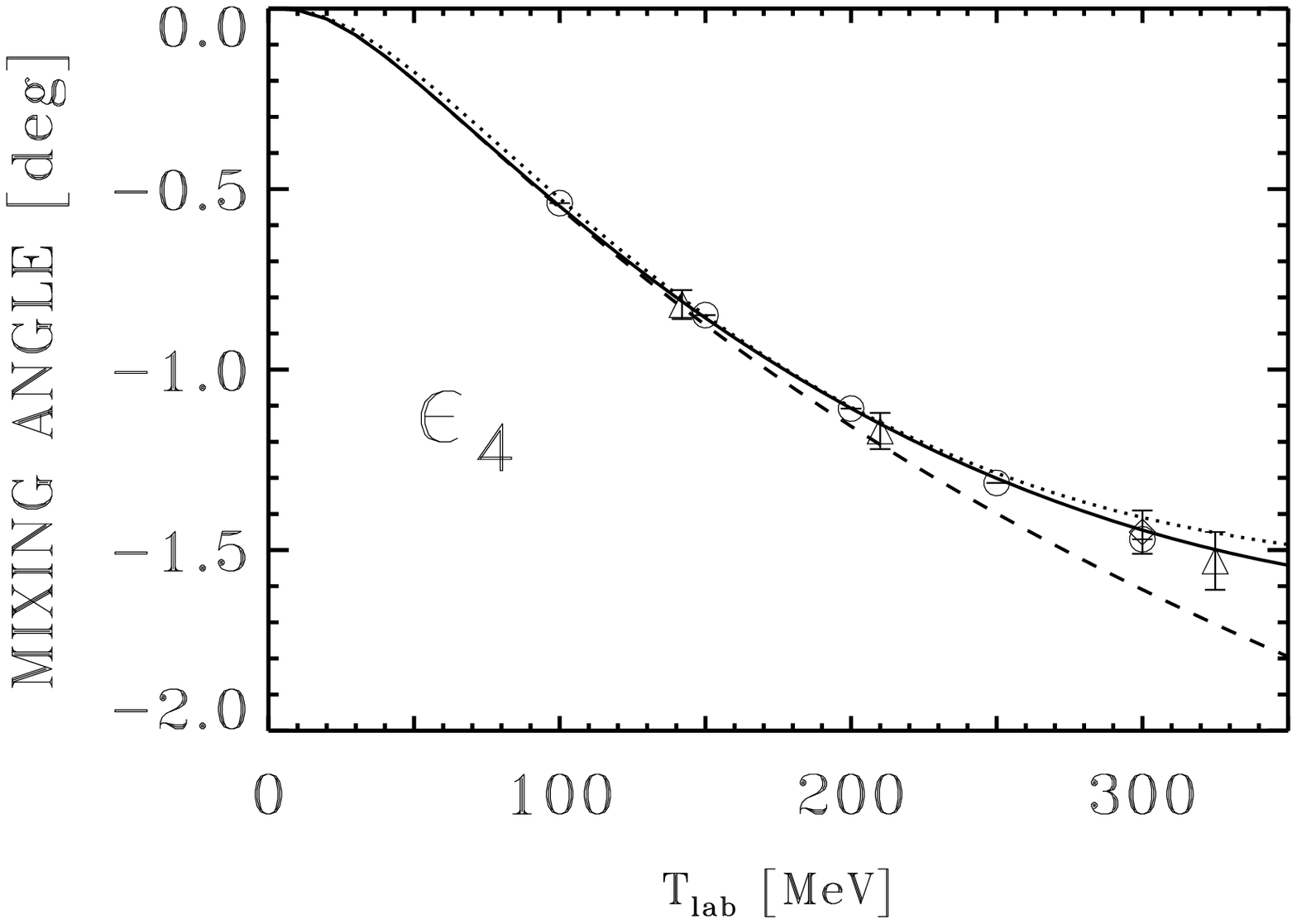,width=80.0mm}}}
\end{picture}
\end{figure}

{\it Fig.6: G-wave  NN phase shifts and mixing angle $\epsilon_4$ versus the
nucleon laboratory kinetic energy $T_{lab}$. For notations see Fig.4.}

\begin{figure}
\unitlength 1mm
\begin{picture}(160,165)
\put(0,110){\makebox{\psfig{file=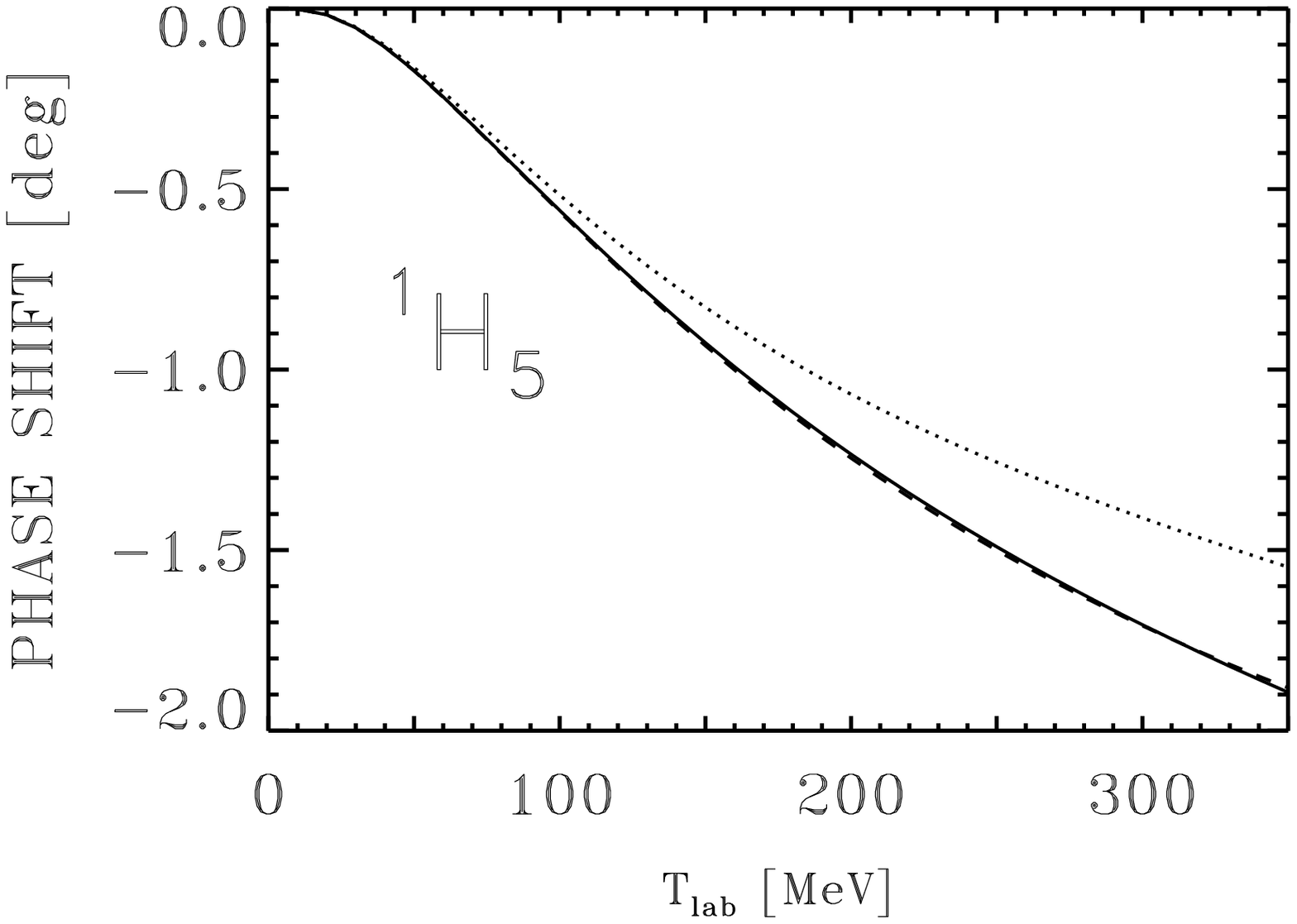,width=80.0mm}}}
\put(80,110){\makebox{\psfig{file=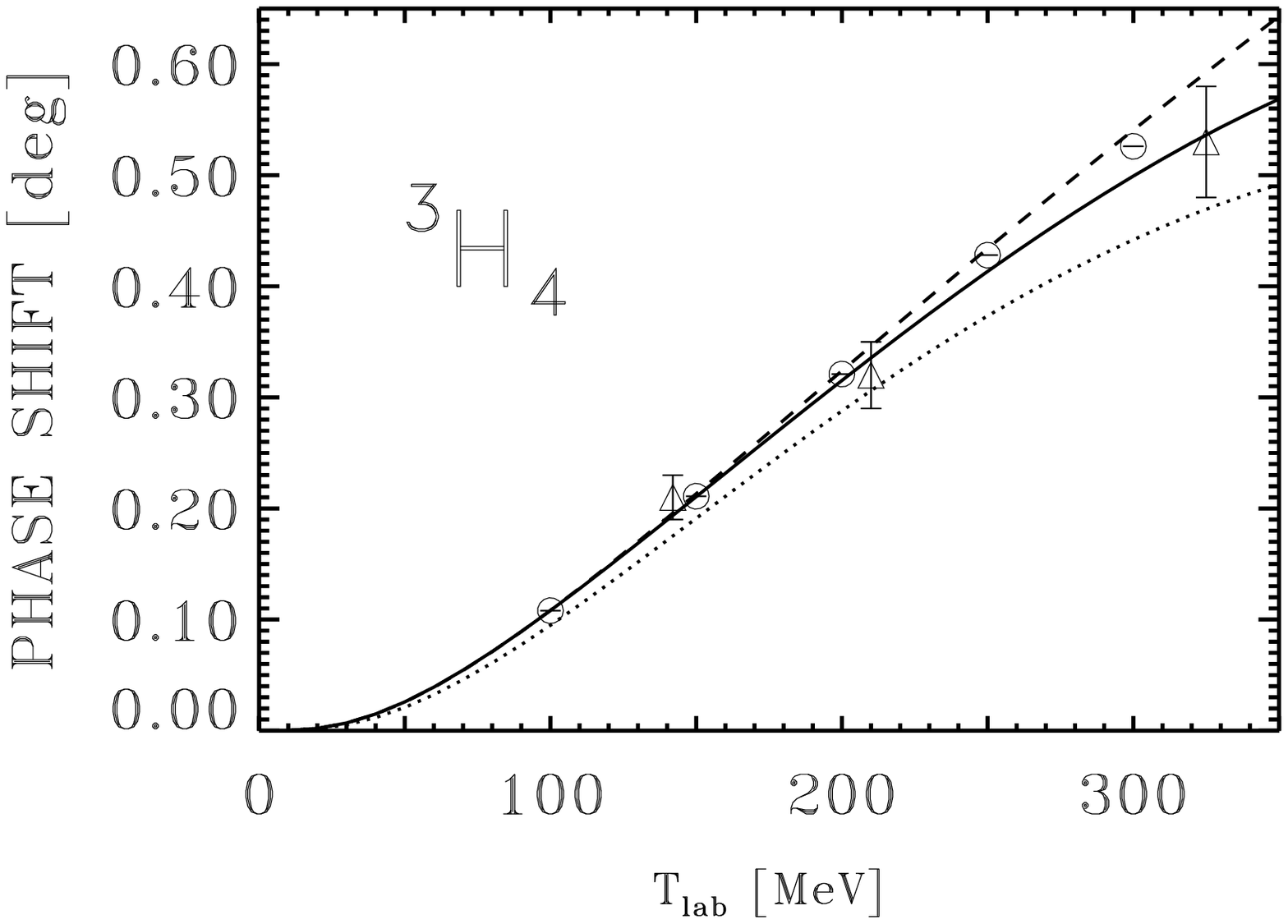,width=80.0mm}}}
\put(0,55){\makebox{\psfig{file=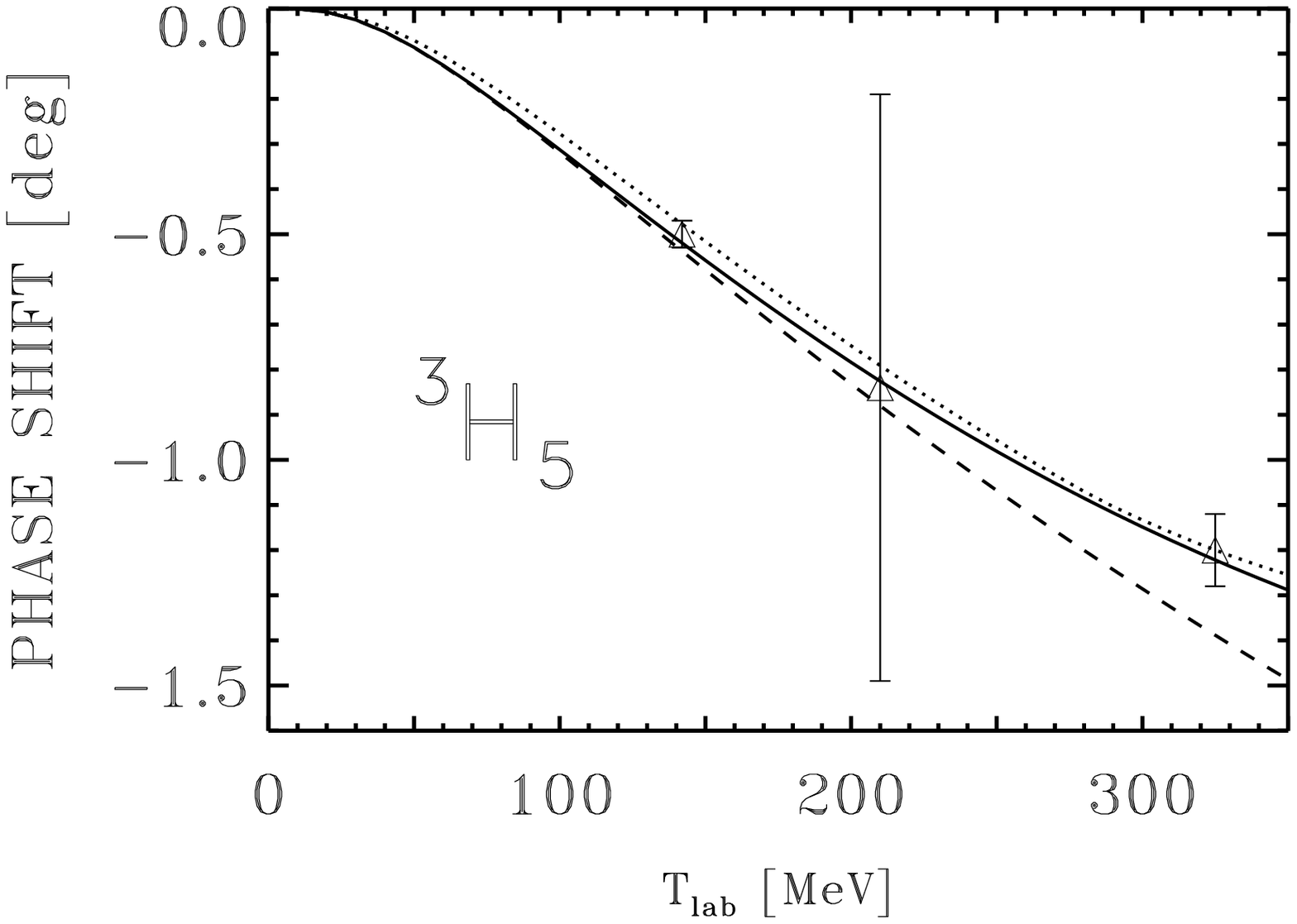,width=80.0mm}}}
\put(80,55){\makebox{\psfig{file=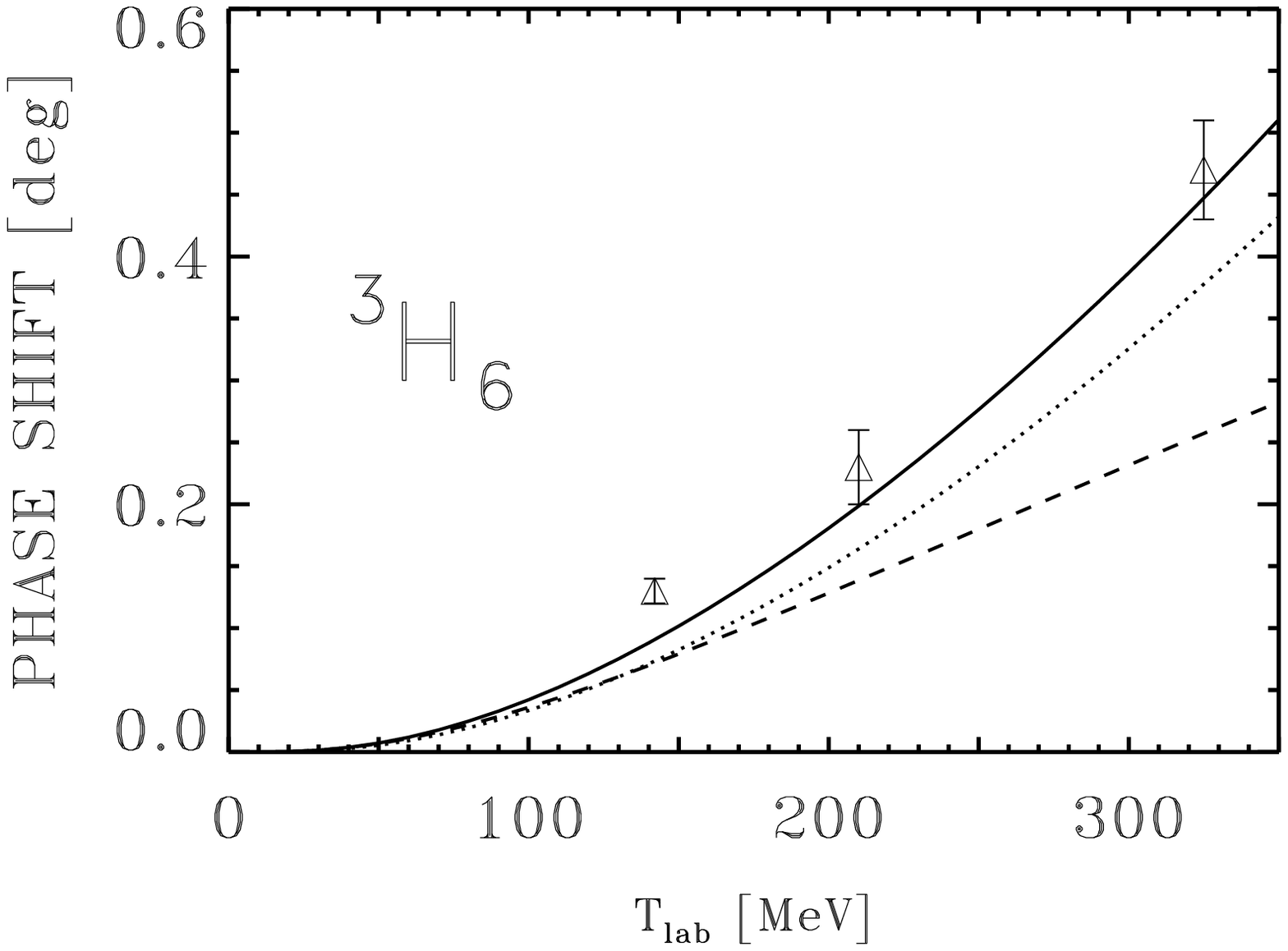,width=80.0mm}}}
\put(40,0){\makebox{\psfig{file=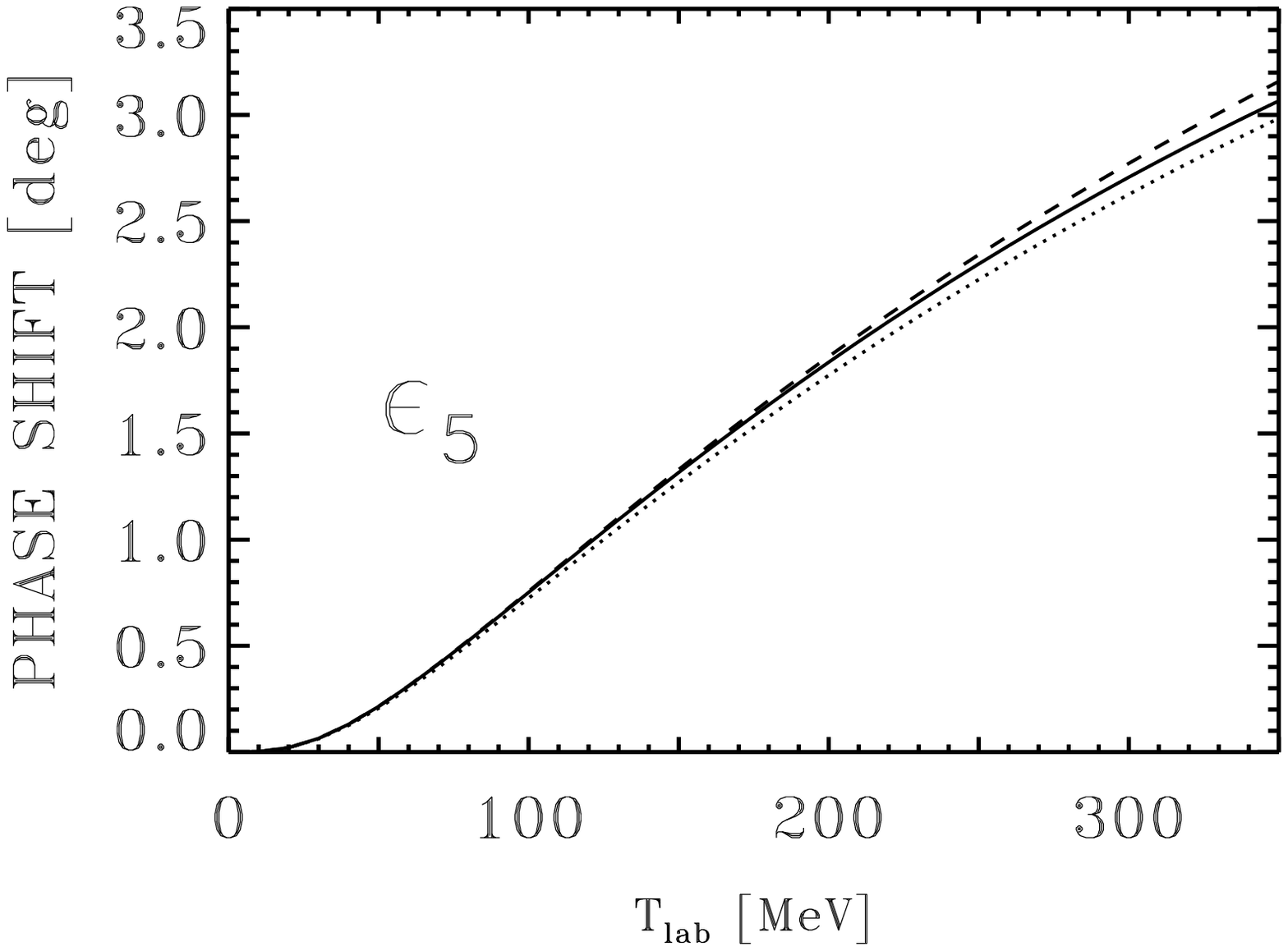,width=80.0mm}}}
\end{picture}
\end{figure}

{\it Fig.7: H-wave  NN phase shifts and mixing angle $\epsilon_5$ versus the
nucleon laboratory kinetic energy $T_{lab}$. For notations see Fig.4.}

\begin{figure}
\unitlength 1mm
\begin{picture}(160,165)
\put(0,110){\makebox{\psfig{file=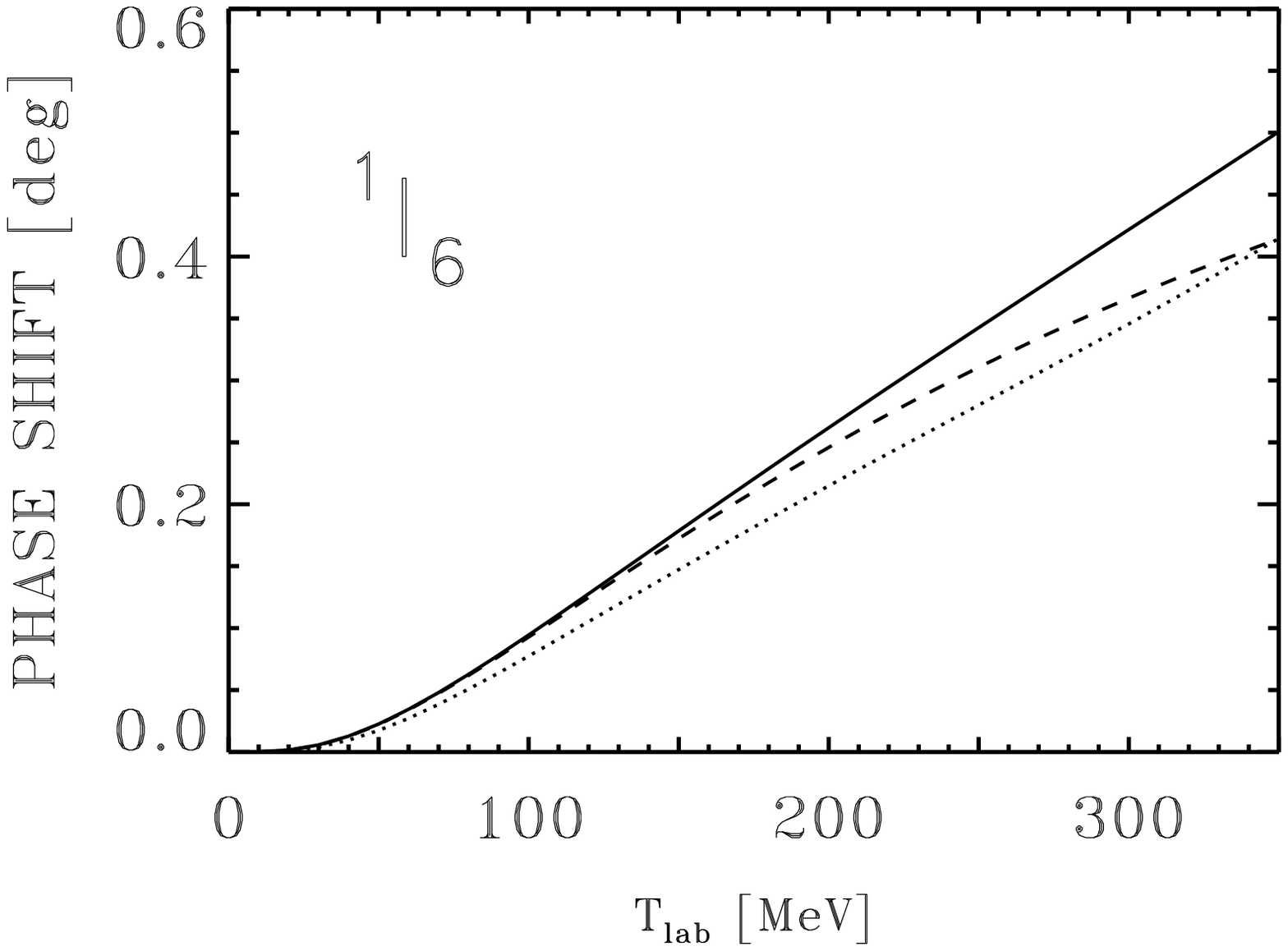,width=80.0mm}}}
\put(80,110){\makebox{\psfig{file=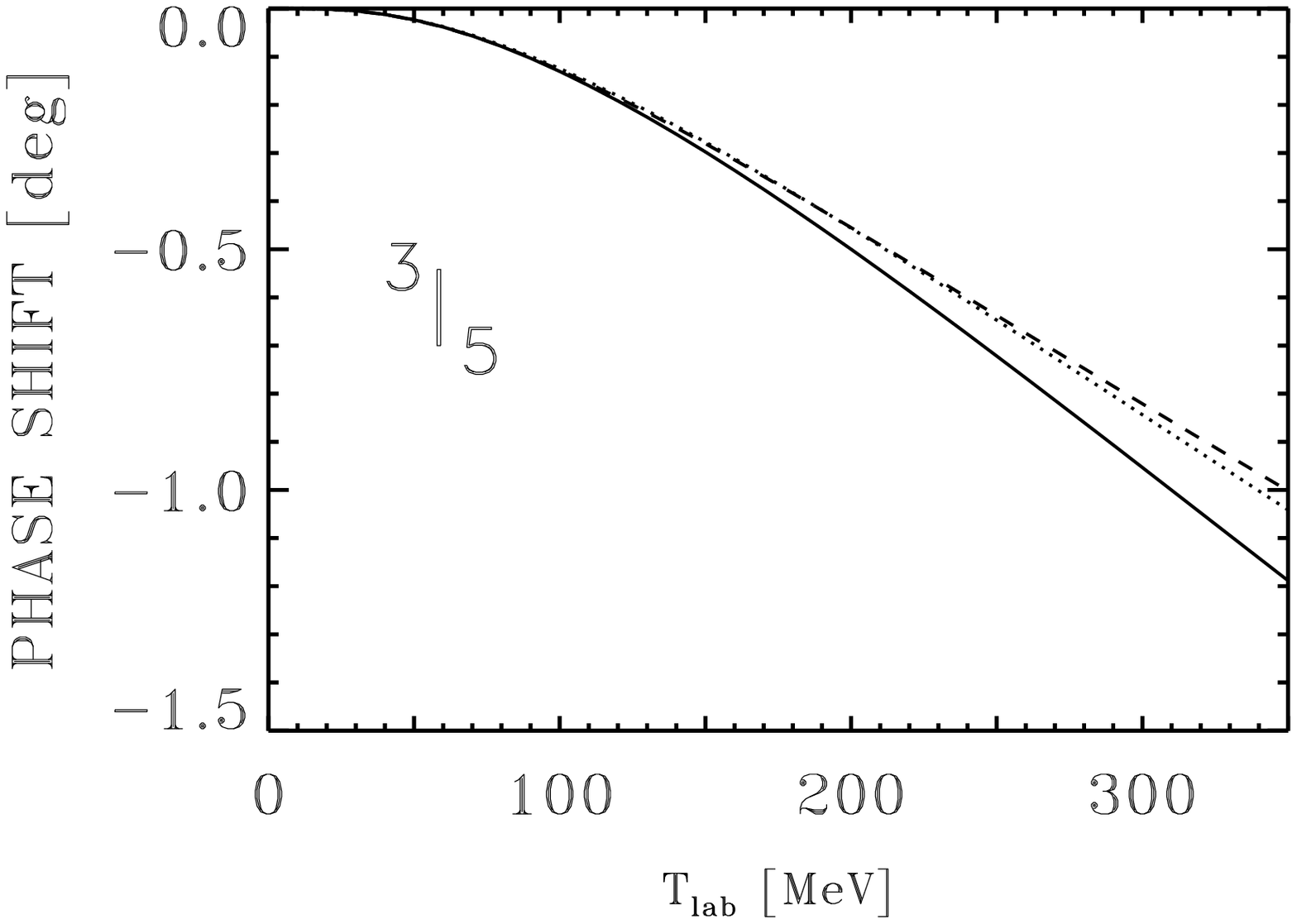,width=80.0mm}}}
\put(0,55){\makebox{\psfig{file=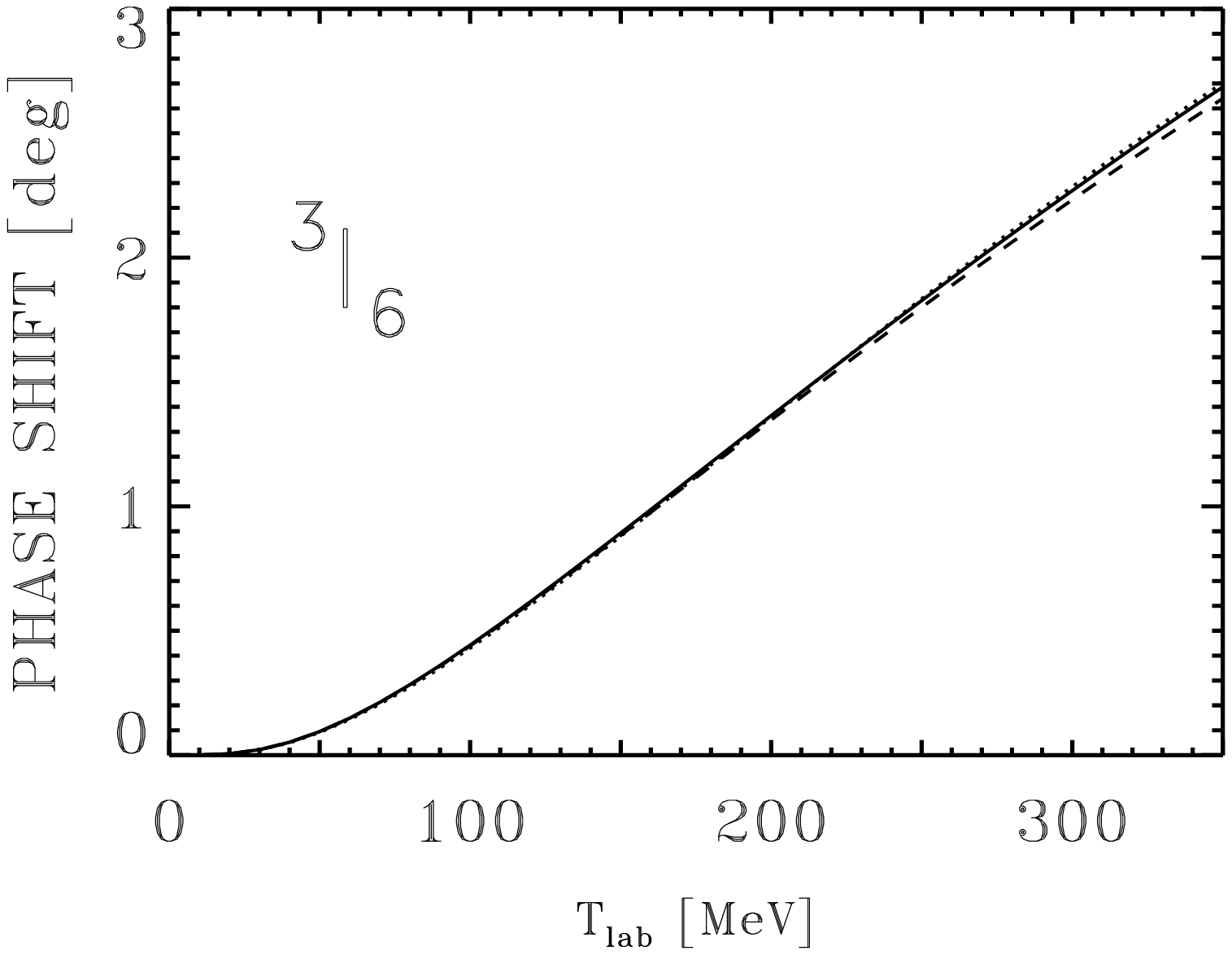,width=80.0mm}}}
\put(80,55){\makebox{\psfig{file=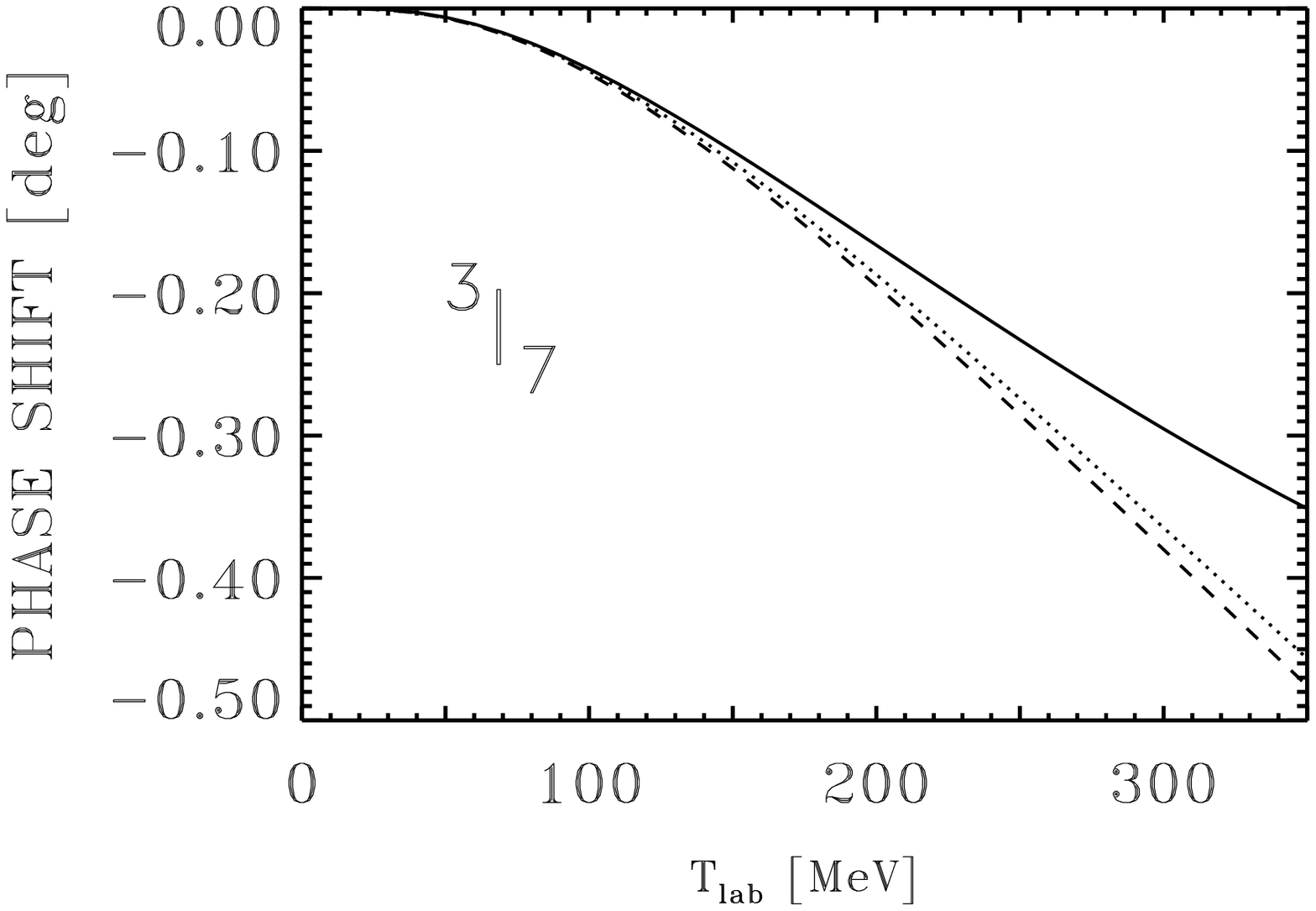,width=80.0mm}}}
\put(40,0){\makebox{\psfig{file=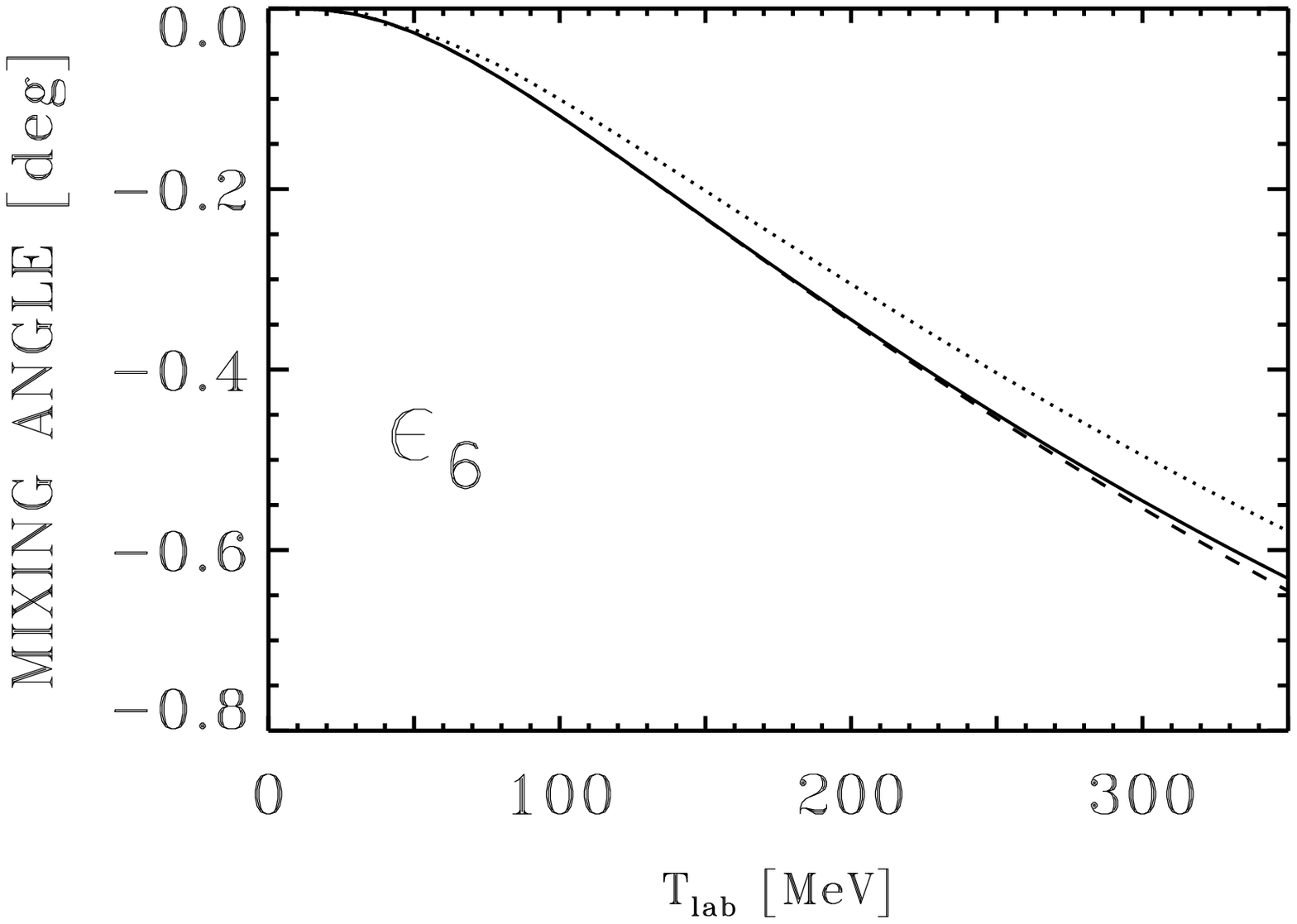,width=80.0mm}}}
\end{picture}
\end{figure}

{\it Fig.8: I-wave  NN phase shifts and mixing angle $\epsilon_6$ versus the
nucleon laboratory kinetic energy $T_{lab}$. For notations see Fig.4.}

\newpage

\begin{figure}
\unitlength 1mm
\begin{picture}(160,55)
\put(0,0){\makebox{\psfig{file=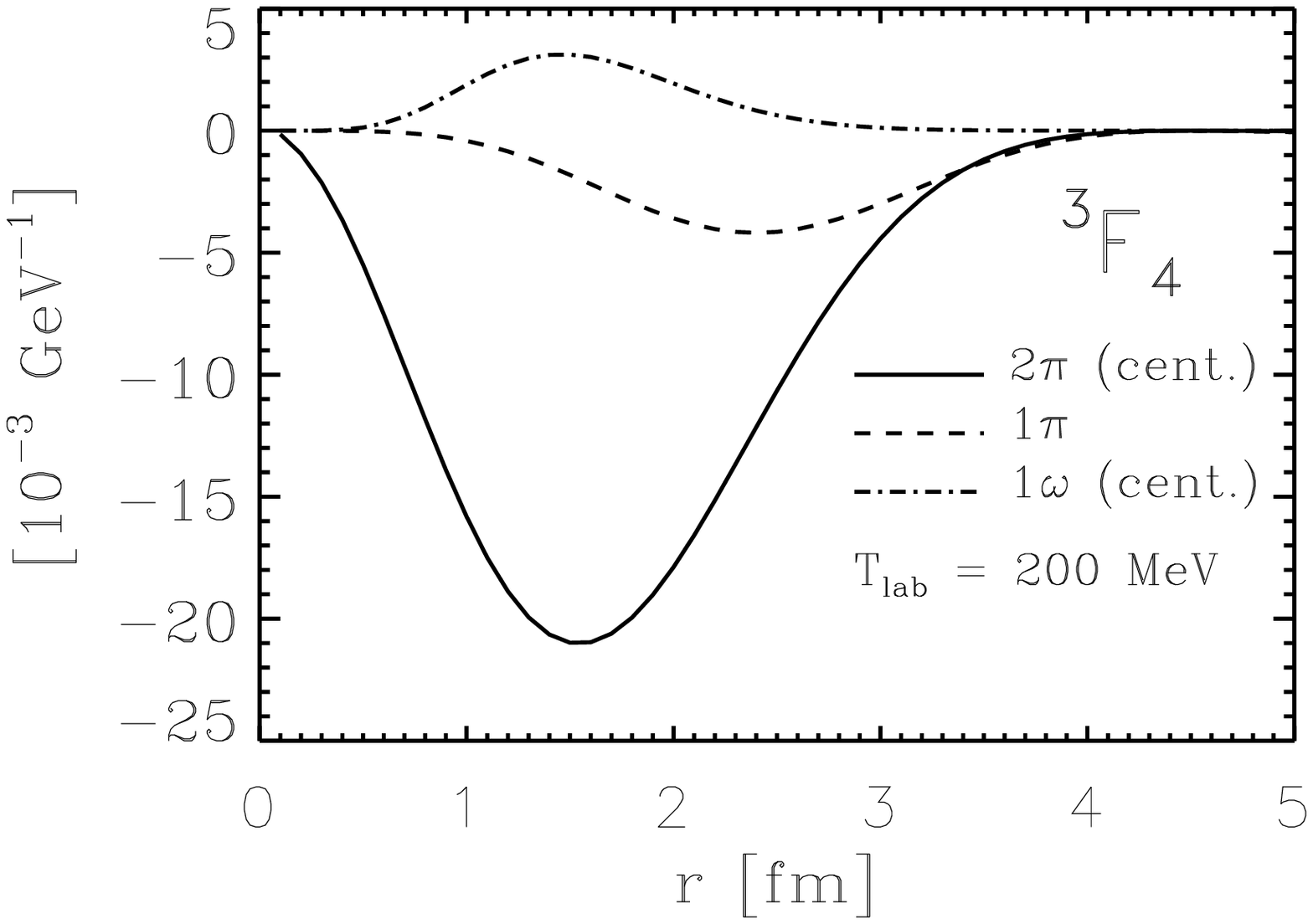,width=80.0mm}}}
\put(80,0){\makebox{\psfig{file=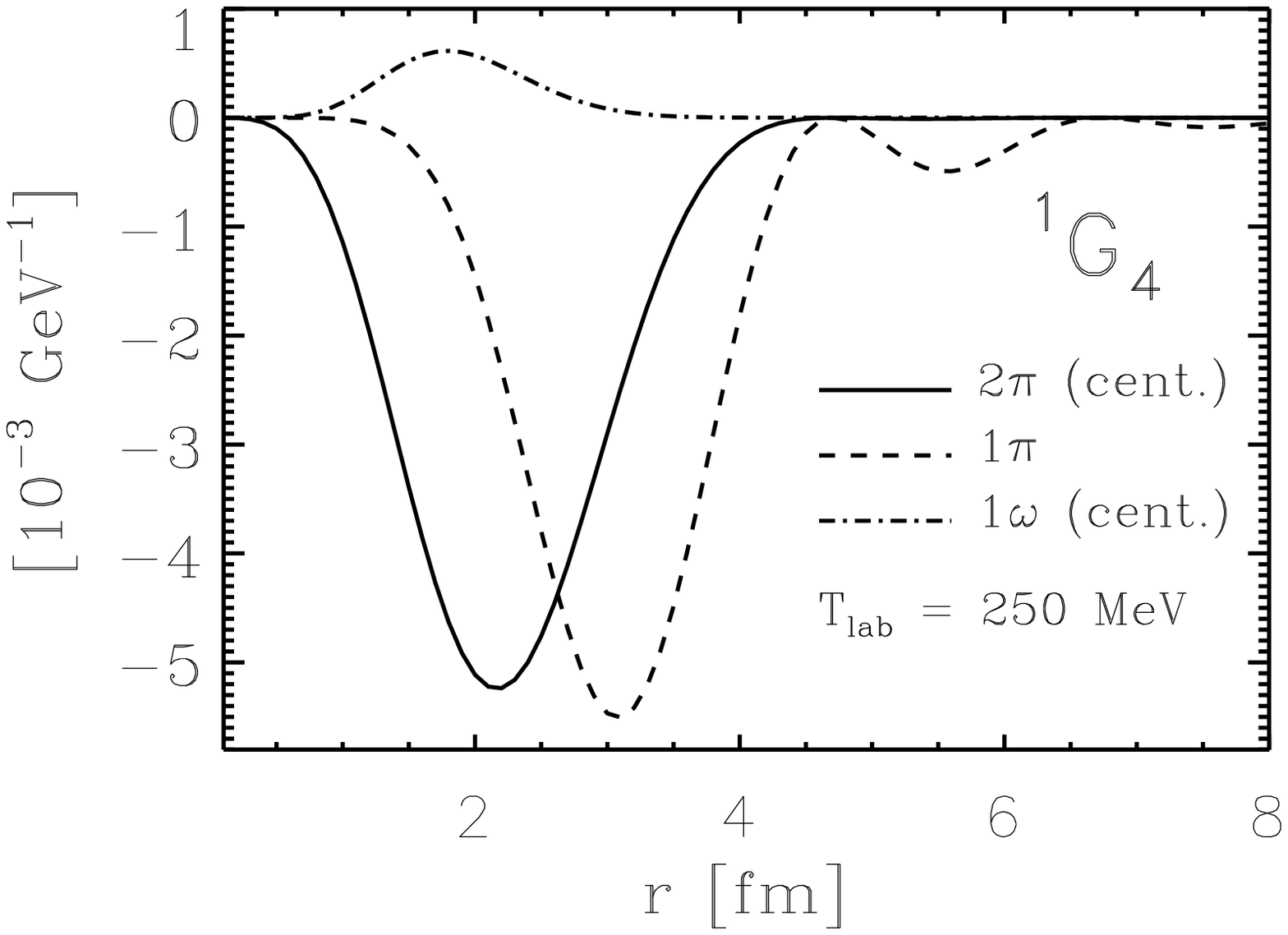,width=80.0mm}}}
\end{picture}
\end{figure}

{\it Fig.9: Examples of  NN interaction densities in coordinate space. Dashed
lines show the $1\pi$-exchange contributions. Full and dashed-dotted curves
give the isoscalar central components of $2\pi$-and $\omega$-exchange, 
respectively.}  

\bigskip
\bigskip

\centerline{\bf Acknowledgement}

\medskip

We thank V. Pandharipande for useful discussions.  

\bigskip

\bigskip

\centerline{ \bf APPENDIX: RELATIVISTIC 1/M-CORRECTION TO $\Delta$-EXCITATION}

\medskip
Here we will collect explicit formulas for the first relativistic correction
proportional to $1/M$ arising from the single and double $\Delta$-excitation
graphs in Fig.1. A convenient way to obtain these corrections is to use 
relativistic propagators and vertices and to perform the $1/M$-expansion inside
the loop integral. For the $\Delta(1232)$ with spin-3/2 this requires the 
Rarita-Schwinger formalism which gives for the $\Delta$-propagator 
\begin{equation} -{i\over 3} {\gamma \cdot k +M_\Delta \over k^2-M_\Delta^2
+i0^+} \bigg\{ 3 g_{\mu\nu} - \gamma_\mu \gamma_\nu - {2 k_\mu k_\nu \over
M_\Delta^2} +{k_\mu \gamma_\nu - k_\nu \gamma_\mu \over M_\Delta}\bigg\} 
\end{equation}
and for the $\Delta \to \pi^a N$ transition vertex
\begin{equation} {3g_A \over 2\sqrt 2 f_\pi} \Big( l^\mu + Z' \gamma \cdot l\,
\gamma^\mu \Big) \, T_a\,\,.  \end{equation} 
$Z'$ is an off-shell parameter lying within in the empirically determined band
$-0.8 < Z' < 0.3$ \cite{benmer}. Again, we omit here additive linear
polynomials $c\,q^2+c'$ in the central amplitudes and additive constants in the
tensor and spin-orbit amplitudes which contain the divergent pieces of the 
diagrams. We will give separately the contributions coming from the three 
classes of $\Delta$-excitation diagrams. We present first the results for 
$Z'=0$ and then display additional $Z'$-dependent terms as far as they show up 
at order $1/M$.  

\medskip

a) $1/M$-correction to triangle graphs with $\Delta$-excitation:

\begin{eqnarray}W_C&=& {g_A^2 \Delta \over 192 \pi^2 M f_\pi^4}\Big\{(
4m_\pi^2+7 q^2)L(q)+3\Sigma\, H(q) +3 (4 \Delta^2 q^2-\Sigma^2) D(q)\Big\}\,\,,
\\ W_{SO}&=& 2W_T = -{2\over q^2}\,W_S= {g_A^2 \Delta \over 128 \pi^2 M f_\pi^4
} \Big\{-2L(q)+(w^2-4 \Delta^2) D(q)\Big\}\,\,.\end{eqnarray} 

b) $1/M$-correction to box graphs with single $\Delta$ excitation:

\begin{eqnarray} V_C&=& {3g_A^4\over 128 \pi^2 M f_\pi^4 \Delta} \Big\{2\big[
 q^4 +2q^2m_\pi^2+2q^2\Delta^2+4m_\pi^4-8m_\pi^6 w^{-2} \big] L(q) \nonumber 
\\ & & +(2m_\pi^2+q^2)^3 \Delta^{-1} \pi A(q) +\Sigma\,
\big[(2m_\pi^2+q^2)^2-4q^2\Delta^2 \big] D(q)\Big\}\,\,,\\  
W_C&=& {g_A^4\over 384 \pi^2 M f_\pi^4 \Delta} \Big\{ 3(2m_\pi^2+q^2)^3 
\Delta^{-1} \pi A(q) +2\Delta^2(8m_\pi^2-q^2) L(q) \nonumber\\& &+3 \Sigma^2\, 
H(q)+ 3\Sigma\, \big[(2m_\pi^2+q^2)^2+4\Delta^2 (q^2- \Delta^2)  \big] D(q 
\Big\}\,\,,
\\  V_T&=& -{1\over q^2}\, V_S= {3g_A^4\over 512 \pi^2 M f_\pi^4 \Delta} \Big\{
(2m_\pi^2+q^2)w^2  \Delta^{-1} \pi A(q) \nonumber \\ & &+ \Sigma\,\big[2 L(q) +
(4\Delta^2+w^2 )D(q) \big] \Big\} \,\,, \\ 
W_T&=& -{1\over q^2}\, W_S= {g_A^4\over 512 \pi^2 M f_\pi^4 \Delta} \Big\{
(2m_\pi^2+q^2)w^2  \Delta^{-1} \pi A(q) \nonumber \\ & &+ 2(4\Delta^2+ \Sigma)
L(q) +\Sigma\,(w^2-4\Delta^2)D(q)\Big\} \,\,, \\ 
V_{SO}&=& {3g_A^4\over 256 \pi^2 M f_\pi^4 \Delta} \Big\{-(2m_\pi^2+q^2)w^2  
\Delta^{-1} \pi A(q) \nonumber \\ & &+2\Sigma \,\big[2
L(q) +(4\Delta^2+w^2 )D(q) \big] \Big\} \,\,, \\ 
W_{SO}&=& {g_A^4\over 256 \pi^2 M f_\pi^4 \Delta} \Big\{2 (2m_\pi^2+q^2)w^2  
\Delta^{-1} \pi A(q) \nonumber \\ & &-2\Sigma\, L(q)
+(4\Delta^2+\Sigma) (4\Delta^2-w^2 )D(q) \Big\} \,\,. \end{eqnarray}

Additional contributions for  $Z'\ne 0$:
\begin{eqnarray} V_C&=& -{3g_A^4Z' \over 32 \pi M f_\pi^4} (3Z'+1)(2m_\pi^2
+q^2)^2 A(q)\,\,, \\ W_T&=&-{1\over q^2}\, W_S = {g_A^4Z' \over 64 \pi 
M f_\pi^4} (3Z'+1) w^2 A(q)\,\,. \end{eqnarray}

c) $1/M$-correction to box graphs with double $\Delta$-excitation:

\begin{eqnarray}V_C&=& {g_A^4 \Delta\over 256\pi^2 M f_\pi^4} \Big\{ 
4(5q^2-16m_\pi^2)L(q)-K(q) \nonumber \\ & & + 3\big[ \Delta^{-2}\Sigma^3 
+8(q^4+6q^2m_\pi^2+2q^2\Delta^2 +8m_\pi^4-8m_\pi^2 \Delta^2) \big] D(q) 
\nonumber \\ & & + 3\big[ \Delta^{-2}(2m_\pi^2+q^2)^2 +4(2m_\pi^2+3q^2 -3
\Delta^2)\big] H(q) \Big\}\,\,,  \\ W_C&=& {g_A^4 \Delta\over 1536\pi^2
M f_\pi^4} \Big\{ 8(12\Delta^2 -8m_\pi^2 + q^2) L(q) -K(q)\nonumber \\ & & + 
3\big[16(2\Delta^4 -2 m_\pi^4 +3q^2 \Delta^2-3q^2m_\pi^2- q^4)- \Delta^{-2}
\Sigma^3  \big] D(q) \nonumber \\ & & +3 \big[ 4(6m_\pi^2+5q^2-5
\Delta^2) - \Delta^{-2}(2m_\pi^2+q^2)^2 \big] H(q) \Big\}\,\,,  \\ 
V_{SO}&=& 4V_T = -{4\over q^2}\,V_S ={3g_A^4 \over 
512\pi^2 M f_\pi^4 \Delta} \Big\{8\Delta^2 H(q) \nonumber  \\ & &
-\Sigma\,\big[2L(q)+ (12\Delta^2+w^2) D(q) \big] \Big\}\,\,, \\  
W_{SO}&=& 4W_T = -{4\over q^2}\,W_S ={g_A^4 \over 
1024\pi^2 M f_\pi^4 \Delta} \Big\{2(\Sigma-8\Delta^2)L(q) \nonumber \\ & & +8 
\Delta^2 H(q)  + (8\Delta^2+\Sigma)(w^2-4\Delta^2) D(q)
 \Big\}\,\,.  \end{eqnarray}  

Additional contributions for $Z'\ne 0$:
\begin{eqnarray} V_C&=& {g_A^4 \Delta Z' \over 16 \pi^2 M f_\pi^4} \Big\{
3\Sigma\, [ 2\Delta^2-\Sigma-Z'(2\Delta^2+3\Sigma) ] D(q) \nonumber \\
& & + 2[7m_\pi^2+4q^2-6\Delta^2+Z'(13m_\pi^2+7q^2-6\Delta^2)] L(q)\Big\}\,\,, 
\\ W_T&=&-{1\over q^2}\, W_S = {g_A^4 \Delta Z' \over 128 \pi^2 M f_\pi^4}
(3Z'+1) \Big\{ (w^2-4\Delta^2) D(q)-2 L(q)  \Big\} \,\,. \quad\end{eqnarray}
The abbreviation $K(q)$ stands for 
\begin{eqnarray}
K(q)&=&{48\, \Sigma^3 \over (w^2- 4 \Delta^2)^2} \bigg[ L(q)- 
{w^2 -4\Delta^2  \over 4 \sqrt{\Delta^2-m_\pi^2}}L'(2\sqrt{\Delta^2-m_\pi^2}) 
-L(2 \sqrt{\Delta^2-m_\pi^2}) \bigg]\,\,, \end{eqnarray} 
and we refer to eqs.(15-19) for the definition of the functions $w,\, \Sigma,\,
L(q),\,A(q),\, D(q)$.  Note that to order $1/M$ only the isoscalar central
$V_C$ and the isovector spin-spin/tensor $W_{S,T}$ NN-amplitudes depend on
the uncertain off-shell parameter $Z'$. Some terms given above have a prefactor
$g_A^4/(M f_\pi^4 \Delta^2)$ of dimension mass$^{-7}$. In coordinate space
these terms will lead to a potential with a rather problematic
$r^{-8}$-singularity near the origin. 

In particular, we have calculated here for the first time the spin-orbit
interaction ($V_{SO}, \, W_{SO}$) generated by $2\pi$-exchange with
$\Delta$-excitation. These spin-orbit NN-amplitudes are truly relativistic 
effects (absent in the static limit and independent of $Z'$) and they may be 
of interest in future studies of the NN spin-orbit interaction.

\end{document}